\documentclass[journal]{IEEEtran}
\usepackage{cite}
\usepackage{graphicx}
\graphicspath{{./figures/}}
\usepackage{tabularx}
\usepackage{amsmath}
\PassOptionsToPackage{hyphens}{url}
\usepackage{xurl}
\usepackage{hyperref}

\usepackage{multirow}
\usepackage{booktabs}
\usepackage{url}
\usepackage{xcolor}

\begin{document}

\title{Grid Integration of AI Data Centers: A Critical Review of Energy Storage Solutions}

\author{Sina~Mohammadi,
        Wayne~Wang,
        Marcus~Chen~I~Wada,
        Rouzbeh~Haghighi,
        Ali~Hassan,
        Hualong~Liu,
        Archit~Bhatnagar,
        Ang~Chen,~\IEEEmembership{},
        and~Wencong~Su,~\IEEEmembership{}
\thanks{Sina Mohammadi, Marcus Chen I Wada, Rouzbeh Haghighi, Ali Hassan, Hualong Liu, and Wencong Su are with the Department of Electrical and Computer Engineering, University of Michigan-Dearborn, Dearborn, MI 48128 USA (corresponding authors: Ang Chen and Wencong Su; e-mail: wencong@umich.edu).}
\thanks{Wayne Wang, Archit Bhatnagar, and Ang Chen are with the Department of Computer Science and Engineering, University of Michigan-Ann Arbor, Ann Arbor, MI 48109 USA (e-mail: angchen@umich.edu).}
\vspace{-5mm}
}

\maketitle

\begin{abstract}

Artificial intelligence (AI) is driving unprecedented growth in data center (DC) scale and power demand. AI workloads impose highly dynamic, difficult-to-forecast power profiles on the utility grid, creating reliability and stability challenges that conventional DC architectures are not designed to address. This paper provides a critical review of energy storage systems (ESSs) as the key enabling technology for reliable grid integration of AI DCs. We organize the review around a four-layer hierarchical taxonomy, namely chip-level buffering, rack/server-level ESSs, facility-level uninterruptible power supply (UPS) systems, and grid-scale battery energy storage systems (BESSs), supplemented by non-battery technologies including fuel cells (FCs) and thermal energy storage (TES). Each layer is analyzed with respect to response timescale, power and energy ratings, operational role, integration challenges, and coordination requirements. Key findings include: (i) AI DC load profiles differ fundamentally from traditional loads in their sub-second variability, making conventional ESS dispatch strategies insufficient; (ii) hierarchical, coordinated ESS deployment across all layers is necessary for effective load smoothing and grid support; and (iii) significant gaps remain in simulation tools, degradation modeling, load forecasting, and optimal multi-layer sizing. This review identifies open research challenges and future directions at the intersection of AI computing infrastructure and power system integration.

\end{abstract}


\vspace{-5mm}
\section*{Abbreviations and Symbols}

\noindent\textbf{Abbreviations}

\begin{tabular}{ll}
AI & Artificial Intelligence \\
AWS & Amazon Web Services \\
BBU & Battery Backup Unit \\
BESS & Battery Energy Storage System \\
BMS & Battery Management System \\
BTM & Behind-the-Meter \\
CapEx & Capital expenditures\\
CPU & Central Processing Unit \\
DC & Data Center\\
DMA & Direct Memory Access \\
\end{tabular}

\begin{tabular}{ll}
DoD & Depth of Discharge \\
DVFS & Dynamic Voltage and Frequency Scaling \\
EMS & Energy Management System \\
ESS & Energy Storage System \\
EV & Electric Vehicle \\
FC & Fuel Cell \\
FFR & Fast Frequency Response \\
FTM & Front-the-Meter\\
GFL & Grid-Following \\
GFM & Grid-Forming \\
GiUPS & Grid-interactive Uninterruptible Power Supply \\
GPU & Graphics Processing Unit \\
HBM & High Bandwidth Memory \\
HDD & Hard Disk Drive\\
HPC & High Performance Computing \\
HVRT & High Voltage Ride Through \\
IEC & International Electrotechnical Commission \\
LFP & Lithium Iron Phosphate \\
LLM & Large Language Model \\
LV & Low Voltage \\
LDESSs & Long Duration Energy Storage Systems\\
LVRT & Low Voltage Ride Through \\
MV & Medium Voltage \\
MVDC & Medium Voltage Direct Current \\
OCP & Open Compute Project\\
ORV3 & Open Rack V3\\
PCC & Point of Common Coupling \\
PDU & Power Distribution Unit\\
PEMFCs & Polymer Electrolyte Membrane Fuel Cells \\
PLL & Phase Locked Loop \\
PPA & Power Purchase Agreement \\
PSA & Power Sharing Unit\\
RDMA & Remote Direct Memory Access\\
RES & Renewable Energy Sources \\
$\text{RoCoF}$ & Rate of Change of Frequency\\
RUL & Remaining Useful Life \\
SEI & Solid Electrolyte Interphase \\
SLB & Second Life Battery\\
SLBESS & Second-Life Energy Storage System\\
SoC & State of Charge\\
SoH & State of Health\\
SOFCs & Solid Oxide Fuel Cells\\
SSD & Solid State Disk\\
SST & Solid State Transformer\\
TDCs & Traditional Data Centers\\
TPU & Tensor Processing Unit\\
TUPS & Traditional UPS\\
VRT & Voltage Ride Through \\
\end{tabular}

\vspace{20pt}
\noindent\textbf{Key Symbols}

\begin{tabular}{ll}
$C_{ch,t}$ & Electricity price during charging \\
& at time step $t$ (\$/kWh) \\
$C_{deg}$ & Degradation cost coefficient (\$/cycle) \\
$C_{dis,t}$ & Electricity price during discharging\\
& at time step $t$ (\$/kWh) \\
$P_{ch}$ & Charging power (kW) \\
$P_{dis}$ & Discharging power (kW) \\
$t$ & Discrete time step (hours) \\
$\Delta f$ & Frequency deviation (Hz) \\
\end{tabular}

\section{Introduction}

Artificial Intelligence (AI) has fundamentally altered society’s relationship with technology, spanning applications from simple search tasks to advanced scientific discovery. Large language models (LLMs) have been at the forefront of these high-complexity AI systems, with models such as ChatGPT and Claude becoming increasingly embedded in everyday use across the United States. These large-scale models are trained on trillions of text tokens collected from online sources and are deployed through web-based platforms that can attract millions of users over short periods of time. This surge in demand has prompted AI industry stakeholders to reconsider contemporary data center (DC) designs so that they can support the intensive computational requirements of training, inference, and fine-tuning for increasingly complex models. Energy requirements for AI queries are now nearly an order of magnitude greater than those of Google search, while AI training workloads have been observed to double approximately every 3.4 months \cite{chalamala2025data}. Collectively, these trends highlight the need to rethink data center architectures to sustainably support the rapid scaling of AI workloads.

DCs are dedicated facilities, or collections of facilities, designed to house computational systems, telecommunications equipment, and data storage infrastructure. Energy demand within a DC is unevenly distributed: IT equipment and cooling systems account for the majority of electricity consumption, while auxiliary loads such as lighting and security represent a comparatively small fraction \cite{CharacteristicsandRisksofEmergingLargeLoads}. Traditional data centers (TDCs) have historically operated at total power levels typically below 30 MW \cite{CharacteristicsandRisksofEmergingLargeLoads}, with individual electrical distribution branches commonly limited to a few tens of kilowatts. These facilities are predominantly connected to power distribution networks in densely populated areas and operate under an architectural paradigm that emphasizes redundancy and exceptionally high uptime, making uninterrupted power delivery from the utility grid essential. TDCs are often sited near population centers to minimize latency for end users.

In contrast, AI DCs are driven by compute-intensive workloads, the widespread deployment of power-hungry GPUs, and the use of high-efficiency cooling technologies, all of which substantially increase power requirements. As a result, large-scale AI training facilities are increasingly located in remote or semi-rural regions where land, power, and cooling resources are more accessible and cost-effective. Rack-level power demand in AI DCs can exceed 100 kW, while aggregate facility consumption may scale to hundreds of megawatts or even gigawatt levels \cite{WhatisanAIdatacenter}. Consequently, large-scale AI training facilities often require direct interconnection with transmission networks and face stricter reliability, siting, and environmental constraints than traditional facilities \cite{DataCenterDevelopmentinanAI-DrivenMarket}. Inference-oriented facilities, while potentially smaller in individual footprint, may connect at either the distribution or transmission level depending on aggregate scale and siting.

\begin{figure*}[t]
    \centering
    \includegraphics[width=0.95\textwidth]{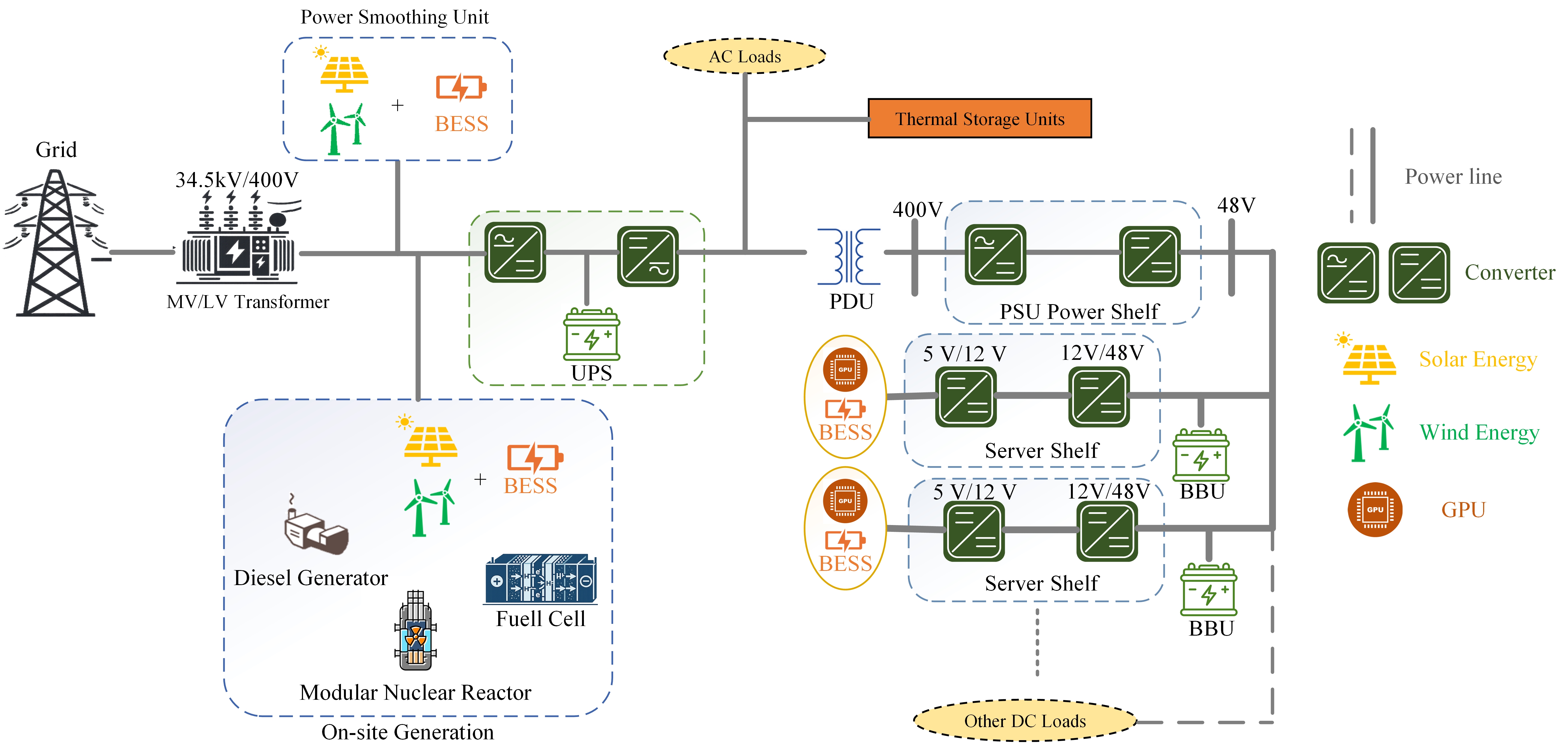}
    \caption{Overall architecture of an AI data center with hierarchical energy storage systems. The figure illustrates the power delivery path from the utility grid through the medium-voltage transformer, UPS/GiUPS, power distribution units, and rack busbars to the IT equipment (GPUs/TPUs).}
    \label{DC overview}
    \vspace{-3mm}
\end{figure*}

Beyond differences in scale and infrastructure, AI DCs also exhibit two distinct load profiles, training and inference, that differ fundamentally from those of TDCs. AI training workloads are characterized by rapid power fluctuations arising from idle periods, peak utilization events, and checkpointing operations, with significant load variations occurring on sub-second timescales \cite{lu2026dynamic,vercellino2026measurement}. In addition, training jobs often produce high-frequency, jitter-like power spikes. In contrast, inference workloads generally lack such rapid transients and exhibit comparatively smoother demand profiles. TDC loads, by comparison, behave more like conventional grid loads and are typically predictable using historical data and standard forecasting methods.

These distinctive load characteristics have important implications for grid operation. AI DCs are connected to the power grid through high-density power converters, which can increase power distortion at the utility level \cite{11369875,xu2026sequential}. In addition, jittery spikes in training jobs may cause sub-frequency mechanical oscillations in synchronous generators, thereby increasing mechanical stress and accelerating corrosion. As a result, harmonic compensation and load smoothing become critical requirements for AI DC operation, both to improve load predictability and to mitigate adverse power-quality impacts on the power system.

Large-scale AI DCs also require the power grid to accommodate higher reserve margins and more proactive resource planning \cite{chen2025electricity}. This is because forecasting their load dynamics is more challenging than for traditional loads, and the uncertainty associated with these facilities is significantly higher. Therefore, additional reserve capacity is needed both to maintain normal grid operation and to support grid expansion when integrating AI DCs.

Moreover, connecting or disconnecting large-scale AI DCs may jeopardize system stability because of high-range frequency fluctuations. These facilities are designed to rapidly transfer to backup power in response to even minor grid disturbances, which can trigger large and sudden load drops at the transmission level. Such disturbances may activate load-frequency relays or force grid operators to implement load-shedding actions. A recent transmission fault in Virginia triggered an abrupt 1,500 MW reduction in system load as multiple DCs transitioned to backup power, leading to significant frequency and voltage deviations that required operator intervention \cite{VirginiaDataCenter}. Events such as this illustrate the limitations of conventional grid planning and operational strategies when accommodating large AI DC loads.

Considering these challenges, this paper investigates the role and impact of energy storage systems (ESSs) in enabling the smooth integration of future large-scale AI DCs into the power system. To the best of our knowledge, there is no comprehensive document specifically focused on the potential support roles of ESSs in the reliable integration of AI DCs with the power grid. In this context, ESSs can serve not only as backup units with limited uptime, but also as enablers of cooling support through thermal storage units. They can be deployed as battery backup units (BBUs) inside IT racks, GPU-level batteries for chip-level load smoothing, and battery energy storage systems (BESSs) for data-center-level load smoothing or co-generation support. ESSs embedded within UPS systems can also interact with the utility grid to provide services such as frequency regulation and peak shaving. In addition, fuel cells (FCs) have the potential to replace diesel backup generation as the technology matures.

To provide a structured perspective on the diverse storage technologies reviewed in this paper, Table~\ref{tab:ess_classification} summarizes the hierarchical ESS classification framework organized by storage layer, typical response timescale, power and energy ratings, primary application role, and the section in which each technology is discussed. This classification is intended to guide readers in comparing technologies across layers and to serve as a roadmap for the remainder of the paper. A more detailed cross-technology comparison, including cost and suitability metrics, is provided in Table~\ref{tab:ess_comparison}.

In contrast to \cite{chen2025electricity,ginzburg2025technical}, this paper focuses specifically on the role of ESSs in enabling reliable grid integration of AI DCs, rather than only characterizing the grid impacts of these large loads. Moreover, our work does not focus on ESS chemistry technologies \cite{rahman2026energy,SAFARI2026120386}, but instead aims to reveal design architectures for the optimal application of ESSs in AI DC grid integration. Therefore, this study serves as an appropriate source for both academic and industry partners seeking the knowledge required to explore new directions for the role of ESSs in future AI DCs. The main contributions of this review paper are summarized as follows:

\begin{itemize}
    \item \textbf{Hierarchical Taxonomy:} We propose a four-layer classification framework organizing ESSs in AI DCs by storage level, response timescale, and application role: (i) chip-level power buffers (sub-millisecond to milliseconds), (ii) rack/server-level BBUs and supercapacitors (milliseconds to seconds), (iii) facility-level GiUPS systems (seconds to minutes), and (iv) grid-scale BESSs (minutes to hours). This taxonomy provides a structured basis for comparing and co-designing ESSs across layers.
    \item \textbf{Grid-Interactive UPS Analysis:} We provide a detailed review of GiUPS operating modes, control architectures, grid-service capabilities (FFR, frequency regulation, VRT, peak shaving), and BESS sizing methodologies, an area that has received limited systematic treatment in prior literature.
    \item \textbf{Comparative Discussion of ESS Technologies:} We critically compare battery-based ESSs, fuel cells, thermal storage, and second-life batteries with respect to their response time, power and energy ratings, application level, advantages, limitations, cost considerations, and suitability for AI DC integration (see Table~\ref{tab:ess_comparison}).
    \item \textbf{Non-Battery ESS Review:} We review the roles of FCs and TES, which are often neglected in AI DC literature, as complementary technologies supporting low-carbon on-site generation and advanced cooling, respectively.
    \item \textbf{Challenges and Research Gaps:} We systematically identify open challenges in simulation and validation, GPU scheduling, degradation modeling, load forecasting, hierarchical coordination, and long-duration storage, and propose specific future research directions.
\end{itemize}

\vspace{-3mm}
\section{AI DC Power Architecture} \label{power architecture}
% \begin{itemize}
%     \item Overview of AI data center power architecture.
%     \item Typical electrical hierarchy: Grid → Power Smoothing Unit → MV/LV Transformer → UPS → Rack Busbar → PSU → Server Shelf.
%     \item Definitions and roles of key units: BESS, UPS, BBU, PSU, PDU, thermal storage, and fuel cells.
%     \item Challenges: Load variability, voltage deviations, harmonic distortion, and fast transients in GPU clusters.
%     \item Need for grid-interactive and multi-layer storage coordination.
% \end{itemize}

The AI DC infrastructure comprises different sections, including IT equipment, power distribution networks, cooling systems, and miscellaneous units such as lighting and security. In this section, we provide a brief introduction to the AI DC components and the ESS units in its structure. The overall structure of the AI DC with all ESSs is provided in Fig. \ref{DC overview}.

\begin{table*}[t]
\centering
\caption{Hierarchical ESS Classification Framework for AI Data Centers}
\label{tab:ess_classification}
\scriptsize
\begin{tabular}{llllll}
\toprule
\textbf{Layer} & \textbf{Technology} & \textbf{Response Timescale} & \textbf{Typical Rating} & \textbf{Primary Role} & \textbf{Section} \\
\midrule
Chip-level & On-chip capacitors, HBM buffers & Sub-ms to ms & W--kW & GPU power spike smoothing & \ref{rack} \\
\midrule
Rack/Server-level & BBUs, supercapacitors, rack BESS & ms to seconds & kW--tens of kW & Server-level load buffering, BBU backup & \ref{rack} \\
\midrule
\multirow{2}{*}{Facility-level} & Traditional UPS (TUPS) & Cycles ($<$20 ms) & 100s kW--MW & Load protection, ride-through & \ref{UPS} \\
                                 & Grid-interactive UPS (GiUPS) & 0.5--10 s (FFR) & 100s kW--MW & FFR, VRT, peak shaving, frequency reg. & \ref{UPS} \\
\midrule
\multirow{2}{*}{Grid-scale} & BTM BESS & Minutes--hours & MW--100s MW & Peak shaving, RES integration, backup & \ref{BESS} \\
                             & FTM BESS & Minutes--hours & MW--GW & Grid ancillary services, reserve & \ref{BESS} \\
\midrule
Backup generation & Fuel Cell (FC) & Seconds--minutes & 100s kW--MW & Low-carbon backup, co-generation & \ref{FC} \\
\midrule
Cooling support & Thermal Energy Storage (TES) & Minutes--hours & MW thermal & Cooling load shifting, RES support & \ref{TE} \\
\midrule
Cost-optimized & SLBESS & Seconds--hours & MW--100s MW & BTM/FTM backup, cost-effective BESS & \ref{SLBESS} \\
\bottomrule
\end{tabular}
\end{table*}

% \vspace{-5mm}
\subsection{IT equipment}
The IT equipment itself comprises high-performance computational electronic chips such as GPUs, TPUs, and CPUs. It also includes advanced storage architectures, such as solid-state storage units, for storing required information with very high speed and efficiency. In addition, it considers resilient and secure networking among all processing units in AI DCs. High-speed, low-latency communication among AI accelerators is primarily achieved through Remote Direct Memory Access (RDMA) fabrics allowing direct memory-to-memory data transfers between nodes without CPU involvement. Dominant implementations include RoCEv2/InfiniBand across racks in training clusters~\cite{alibaba-hpn} and direct GPU-to-GPU communication within racks via NVLink~\cite{GPU-to-GPUCommunication:UnlockingParallelismBeyondtheCore}.

\subsubsection{Computing Units}
% General overview of the chips

The computing unit consumes half of the required electricity of an AI DC \cite{NERC2025LargeLoads}. NVIDIA GPUs such as H100, B200, or Blackwell Ultra have been used for training jobs in hyperscale AI DCs. For instance, the B200 utilizes 208 billion transistors that can be parallelized through very fast communication settings such as NVLink \cite{AnOverviewofPopularNVIDIAGPUs}. In addition, new GPU designs such as the GB300 NVL72 enable chip-level power smoothing using ESS \cite{HowNewGB300NVL72FeaturesProvideSteadyPowerforAI}. 
Beyond NVIDIA, major cloud providers have developed custom AI accelerators optimized for their infrastructure: Google's Tensor Processing Units (TPUs)~\cite{AccelerateAIInferencewithGoogleCloudTPUsandGPUs} are used to train and serve models such as Gemini, while AWS Trainium and Inferentia chips provide training and inference acceleration, respectively, enabling cloud-scale AI services.

AMD has also entered the AI accelerator market with its Instinct GPU series (e.g., MI300X), which offers competitive performance-per-watt for LLM training and inference workloads. These GPUs integrate CPU and GPU compute with high-bandwidth memory (HBM) in a unified package, positioning it as a direct competitor in hyperscale AI DCs. In section \ref{rack}, the role of ESSs for power smoothing will be discussed in more detail.

% In addition to NVIDIA and its GPUs, Intel and AMD have launched new CPUs that are able to support high computational tasks such as AI workloads in data centers. For instance, the Intel Xeon 6 processor and AMD EPYC CPUs are designed and developed with higher capacities and efficiency compared to older CPU versions \cite{IntelUnveilsLeadershipAIandNetworkingSolutionswithXeon6Processors}. As AMD claims, the new CPU design can be utilized in AI DCs without using GPUs, and it can extend cluster efficiency by 20% \cite{TheAMDAdvantageforAIandDataCenters}.

% In this paper we investigate the role of chip-level energy storage units in addressing the load smoothing of GPUs power profiles. The types of energy storage units are discussed for damping the power large spikes and also jittery fluctuations during the normal training workloads. Also, the coordination of chip-level storage units with BBUs are discussed. 

\subsubsection{Data Storage Units}
AI training workloads place extreme demands on storage infrastructure, requiring high-throughput, low-latency access to datasets that can reach petabyte scale. NVMe SSDs have become the dominant storage medium in modern AI DCs, replacing both HDDs and legacy SATA SSDs due to their superior bandwidth and latency characteristics. HDDs, 
while still present in archival or cold-storage tiers of some traditional facilities, are largely absent from the hot data path in AI training environments, where even brief I/O stalls can idle 
expensive GPU clusters \cite{WhatchangesinstoragewillAIdrive?}.

The prevailing architectural trend in hyperscale AI DCs is \textit{disaggregated storage}, in which compute (GPU servers) and storage are decoupled into independent resource pools interconnected via high-speed RDMA fabrics. Unlike traditional server-attached storage, disaggregation allows compute and storage capacity to scale independently, improves overall resource utilization, and enables storage to be shared across multiple training jobs simultaneously \cite{alibaba-hpn}. In practice, GPU servers access remote NVMe devices over RDMA-capable networks using protocols such as NVMe over Fabrics (NVMe-oF), achieving latencies that approach those of locally-attached storage. NVIDIA's GPUDirect Storage extends this further by enabling direct DMA transfers between NVMe storage devices and GPU memory, bypassing the CPU and host memory entirely and reducing both latency and CPU overhead for I/O-bound training 
workloads \cite{GPUDirectStorage:ADirectPathBetweenStorageandGPUMemory}.

\subsubsection{Networking}
AI DCs aggregate thousands of GPU accelerators into tightly coupled 
parallel training clusters, placing extraordinary demands on the 
underlying network fabric. Unlike TDCs, where network traffic 
patterns are largely client-server in nature, AI training workloads 
generate dense all-to-all communication patterns during gradient 
synchronization across distributed model replicas. This requires 
network fabrics that deliver high bandwidth, ultra-low latency, and 
minimal congestion at scale \cite{alibaba-hpn}.

The foundational transport mechanism in modern AI DC networks is RDMA, which enables direct 
memory-to-memory transfers between nodes without CPU involvement, reducing both latency and CPU overhead. While InfiniBand has  historically been used in HPC settings and remains present in some training clusters, the industry has broadly converged on RoCEv2 (RDMA over Converged Ethernet)~\cite{meta_roce} as the fabric of choice at 
hyperscale, offering comparable performance over commodity Ethernet switching infrastructure at lower cost and with greater operational 
flexibility \cite{alibaba-hpn, NetworkingforDataCentersandtheEraofAI}. 
NVIDIA's Spectrum-X platform is specifically engineered to optimize RoCEv2 performance for AI workloads, addressing congestion control and load balancing challenges that arise in large-scale training deployments \cite{NetworkingforDataCentersandtheEraofAI}.

Within a single node or rack, inter-GPU communication bypasses the network fabric entirely through NVIDIA NVLink, a high-bandwidth direct interconnect that enables GPU-to-GPU data transfers at bandwidths far exceeding those achievable over any network 
interface \cite{GPU-to-GPUCommunication:UnlockingParallelismBeyondtheCore}. 
Beyond a single rack, scale-out communication relies on the RDMA fabric described above. 

% \subsubsection{Networking}
% AI DCs collect thousands of AI accelerators (GPUs, TPUs, or CPUs) to enable high-intensity parallel processing. This large integration requires high-speed, low-latency networking systems, including NVIDIA InfiniBand NDR and high-speed Ethernet \cite{AIDataCenterInfrastructureEssentials:BuildingTopologiestoSupportAIandMLWorkloads}. In addition, the NVIDIA Spectrum-X networking platform can enhance Ethernet capabilities in AI DCs \cite{NetworkingforDataCentersandtheEraofAI}.

% TDCs use fiber optics for external communications and copper wires for server-level communications. In AI DCs, there is a need to use fiber optics inside the IT equipment to enable high-speed communication among AI accelerators. Multimode fiber cables have lower installation costs compared to single-mode cables \cite{FiberOpticCableTypes–MultimodeandSingleMode}. GPU-to-GPU communication without the need for a CPU is also an alternative method for enhancing data-sharing speed among GPUs, which is now available through NVLink \cite{GPU-to-GPUCommunication:UnlockingParallelismBeyondtheCore}.

% \vspace{-3mm}
\subsection{Power Delivery Systems}
The power delivery systems in an AI DC contain the external utility grid, internal power distribution, UPS, on-site backup generation, and the associated control systems.

\subsubsection{External Utility Grid}
TDCs, with power consumption on the order of a few megawatts, can typically be directly connected to distribution grids at voltage levels such as 34.5 kV or 13.2 kV. However, hyperscale AI DCs, requiring more than 100 MW of power, should be connected to transmission grids to ensure reliable and stable grid operation \cite{CharacteristicsandRisksofEmergingLargeLoads}. However, there is no standardized voltage level across utilities for integrating large loads such as AI DCs. The utility power is then transformed to a lower voltage level through a medium-voltage to low-voltage distribution transformer, typically 480 V, to feed the internal power distribution within the DC. This transformer also electrically isolates the AI DC from the grid. The 480 V AC is then distributed among the various components of the AI DC, including the UPS system, power distribution units (PDUs), cooling system, and IT equipment.

\subsubsection{UPS system}
The UPS system is comprised of an AC-to-DC power converter, a battery storage unit, and a DC-to-AC power converter. Power is stored in the storage unit under normal conditions, and the stored energy is used during emergency conditions when there is a loss of utility power or rapid fluctuations that jeopardize the normal operation of IT loads. In TDCs, UPS operation is typically limited to a few seconds, after which on-site generation units take responsibility for powering the DC in the absence of utility power for several hours.

The UPS requirements for AI DCs differ significantly from those of TDCs. The UPS system for AI DCs should operate over different ranges, including short-, medium-, and long-duration operation. Medium- and long-duration operation provides more time for on-site generation activation or, in some cases, can eliminate the need for separate backup units. GiUPS systems with high power density and large storage capacity can also be implemented to enable grid-support functions, including peak shaving or frequency regulation scenarios. The UPS systems in modern AI DCs will be discussed in detail later in Section \ref{UPS}.

\subsubsection{Backup On-Site Generation}
Backup generation units are utilized in DCs to provide electrical power under emergency conditions when utility power is unavailable. Backup generation is activated a few seconds after a utility power interruption, following the UPS response. Diesel backup generators are used in TDCs because they are combustion-based units that can provide a fast response; however, they are not environmentally friendly. RESs with appropriate BESSs can also be utilized for backup generation, as discussed in more detail in Section \ref{BESS}. However, power density, implementation cost, and the required space for installing PV panels or wind turbines are challenging issues in this context \cite{10255526}. Modular nuclear reactors have also recently been introduced for AI DCs, offering MW-level power generation \cite{AdvantagesandChallengesofNuclear-PoweredDataCenters}. The technology for this type of equipment is still immature, implementation costs are high, and spent nuclear fuel must be stored properly. In addition, FCs may be better options for backup generation, as they can provide high power with fast response. Moreover, the development of hydrogen-based FCs can enable clean generation units within AI DCs \cite{11230048}. The role of FCs in backup generation for future AI DCs will be discussed thoroughly in Section \ref{FC}.

% \vspace{-3mm}
\subsection{Cooling Systems}
The IT equipment inside a DC produces heat as a result of electrical losses within electronic chips. As the computational power and power density of computing units increase, the associated losses and heat generation also increase. In AI DCs, due to the presence of high-intensity AI accelerators, advanced cooling systems are required to enhance IT load performance and optimize power consumption. In TDCs, air is used as the primary medium for cooling IT equipment. Chilled air is injected into server rooms, and the temperature is maintained at a certain level using a central cooling controller. Air-cooling systems are cost-efficient, reliable, and agile. They can be easily extended to larger facilities, and the required hardware can be modified quickly. However, the efficiency of air-cooling systems is low for effectively cooling power-hungry AI accelerators.

Liquid cooling is a more advanced cooling system developed for AI DCs. By utilizing liquid cooling systems, cooling can be provided at the rack level or even at the chip level. Liquid cooling offers better heat transfer, requires less energy and space, and is more efficient compared to air-cooling systems. It also eliminates the need for large mechanical fans, which are less reliable. Direct-to-chip cooling circulates a dielectric liquid through cold plates into chips and processing units and can efficiently dissipate heat \cite{Understandingdirect-to-chipcoolinginHPCinfrastructure:Adeepdiveintoliquidcooling}. Hybrid cooling systems combine the agility of air cooling with the efficiency of liquid cooling, making them a suitable option for hyperscale AI DCs where redundancy can be achieved through hybrid solutions.

\begin{table*}[t]
\centering
\scriptsize
\caption{Overview of GiUPS Operating Modes}
\begin{tabular}{lccll}
\hline
\hline
\textbf{Mode} & \textbf{Grid Input} & \textbf{Battery Output} & \textbf{Trigger Condition} & \textbf{Key Feature} \\
\hline
Standard Operation & 100\% & 0\% & Normal operation & Standard double-conversion UPS \\
\hline
Discharge-Full Disconnection & 0\% & 100\% & External command & Fully battery-powered load \\
\hline
Discharge-Partial Disconnection & $<$100\% & Partial (e.g., 25\%) & External command & Mixed grid and battery supply \\
\hline
Recharge & Supply $>$ Load & Battery charging & SOC $<$ 100\%, over-frequency & Power limited (20--25\% of UPS capacity) \\
\hline
Energy Export & Adjustable & Discharge to grid & Regulatory approval & Bi-directional power conversion \\
\hline
\hline
\end{tabular}
\label{GiUPS operation overview}
\vspace{-3mm}
\end{table*}

TE units can also be utilized in parallel with different types of cooling systems \cite{takci2025data}. The TEs absorb chilled air or water and inject it into the IT equipment at specific times. Utilizing TEs can effectively increase efficiency in AI DCs. TEs cool the air or liquid by accessing extra energy from on-site generation units during normal DC operation. The role of TEs as cooling system accelerators in future AI DCs will be discussed thoroughly in Section \ref{TE}.

% \vspace{-3mm}
\subsection{Miscellaneous Units}
Apart from IT equipment, power delivery, and cooling systems, there are other units required for the normal operation of AI DCs. These include the physical infrastructure for housing servers, racks, and wiring systems, as well as the lighting system within the AI DCs. In addition, security and control rooms are also integral units of a AI DC. Miscellaneous units in an AI DC consume significantly less electricity than the other sections.

{\begin{figure}[!b]
\centering
\includegraphics[width=0.95\linewidth]{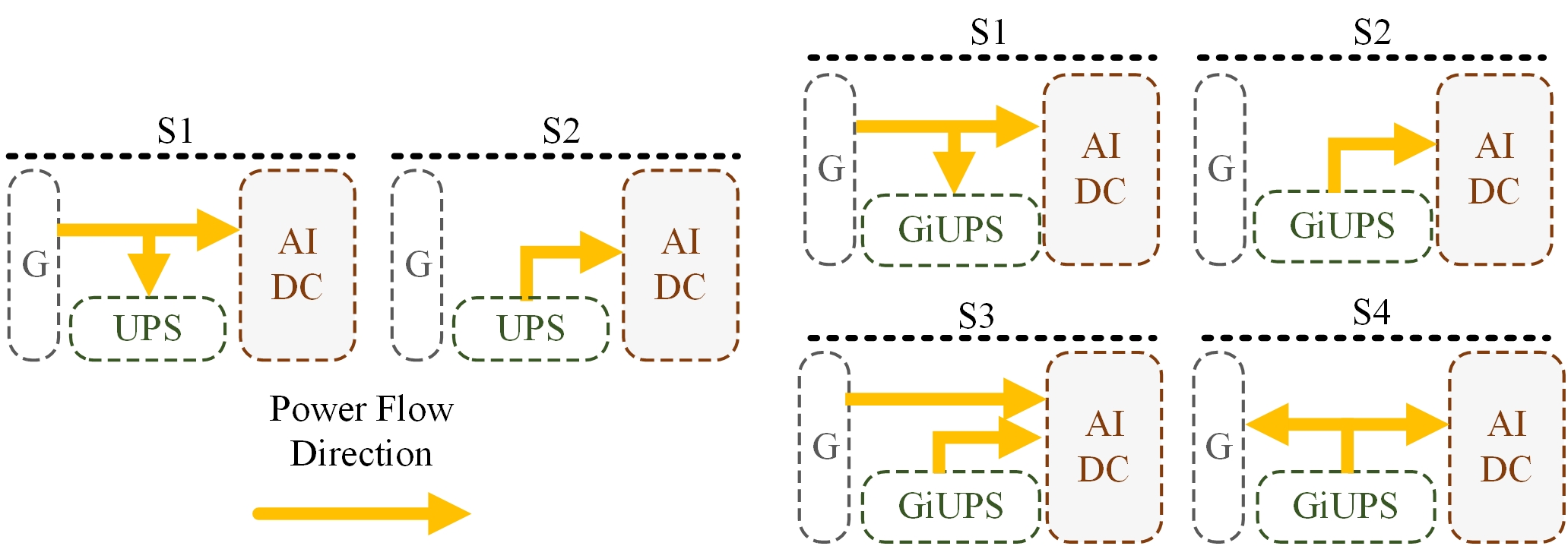}
\caption{UPS and GiUPS operation modes}
\label{UPS_GiUPS operation}
\vspace{-3mm}
\end{figure}}

\section{UPS Systems in AI DCs} \label{UPS}

\subsection{Traditional UPS (TUPS) Operation and Limitations}
UPS systems are fundamental components in AI DCs, ensuring continuous operation of critical loads during grid disturbances or outages. Conventional UPS systems~\cite{karpati2012uninterruptible,fawaz2019minimizing,wang2025coordinated} typically operate in standby or online modes. In standby mode, the UPS remains bypassed during normal operation and transfers to battery-supported inverter operation only when a disturbance is detected. In online mode, the rectifier and inverter continuously process power, providing isolation from grid disturbances and maintaining tight voltage and frequency regulation at the critical bus (see Fig. \ref{UPS_GiUPS operation}). In both modes, the UPS interacts with the grid only through battery charging and synchronization. The UPS is therefore treated as a passive protective device.

% \vspace{-3mm}
\subsection{GiUPS systems in AI DCs}
TUPS architectures were designed primarily for load protection with minimal interaction with the utility grid. This paradigm is increasingly insufficient for AI DCs. First, AI workloads introduce rapid load variations that propagate upstream and cause voltage flicker, transformer thermal stress, and feeder congestion~\cite{10255103}. These effects are amplified when multiple AI racks operate in synchrony, which is common in distributed training. Second, the growing penetration of renewable energy reduces system inertia and increases the rate of change of frequency. TUPS systems do not contribute to frequency support beyond simple ride-through, which leaves a large amount of inverter capacity unused during normal operation~\cite{colangelo2025ai}. Third, the energy stored in UPS batteries remains idle for more than 99 percent of their lifetime, representing a significant opportunity cost. These limitations motivate the development of UPS architectures capable of dynamic grid interaction.

% However, due to the challenges raised by AI DCs discussed earlier, the industry has begun to explore GiUPS systems. These systems preserve load continuity while also providing grid services such as frequency regulation and peak shaving~\cite{paananen2023grid}.

% \subsubsection{GiUPS Working Mechanism}
GiUPS systems enhance TUPS designs by incorporating bidirectional power converters, advanced digital controllers, and communication interfaces that allow coordination with the utility grid or market operator. These systems can modulate active power in response to grid frequency deviations or dispatch signals, enabling participation in services such as primary frequency regulation, fast frequency response, synthetic inertia, and peak load management~\cite{Vertiv2025GridInteractiveUPS,paananen2021grid,MediumVoltageUPS}. Unlike TUPS systems, which operate independently of grid conditions except during outages, GiUPS systems continuously monitor grid conditions and adjust their output accordingly. 

% The GiUPS detects load transients by continuously monitoring the incoming utility frequency. This is achieved through sub-cycle frequency measurement using a digital synchronous reference frame phase-locked loop (PLL) implementation. Based on the observed frequency deviation (Hz) and the rate of change of frequency (RoCoF, Hz/s), the UPS computes its required power response at an update rate of approximately 5 ms \cite{Eaton2025DataCentersGoodGridCitizen}. 

% The control architecture of GiUPS typically includes a multi-layer structure. The inner control loops regulate current and voltage at sub-millisecond timescales. The outer loops implement grid-following or grid-forming behavior depending on the operating mode. A supervisory controller coordinates battery SoC, grid service participation, and critical load protection. This layered structure allows the UPS to respond to frequency deviations within tens of milliseconds, which is significantly faster than most synchronous generators and comparable to dedicated BESSs~\cite{takci2025data}.

GiUPS can provide fast frequency response (FFR) within 0.5 to 10 seconds after a grid event. It can operate in a dynamic frequency response mode. In this mode, the injected battery power is adjusted in real time based on frequency regulation needs. The UPS can also operate in a static frequency response mode. In this case, a fixed amount of power is injected after a grid event such as a fault or large load switching \cite{MaximizeRevenues}.

\subsubsection{Operational Modes of GiUPS}

GiUPS can operate in several modes depending on the grid conditions and external control signals, including \textit{standard operation}, \textit{discharge modes} (full and partial disconnection), \textit{recharge mode}, and \textit{energy export (bi-directional) mode}~\cite{Vertiv2025GridInteractiveUPS,eaton_microsoft_energy_aware_ups}. When a frequency variation is detected by an external controller, the GiUPS adjusts its operation to provide both positive and negative regulation by charging or discharging its batteries within operational limits. Table \ref{GiUPS operation overview}, shows an overview of GiUPS operation modes.

Under normal conditions, the GiUPS receives 100\% of the input power from the utility grid through the rectifier, delivering it to the load (see S1 in Fig. \ref{UPS_GiUPS operation}). In this mode, the GiUPS functions as a standard double-conversion system, with the batteries maintained in standby and not actively supplying power. This operation is referred to as the \textit{standard UPS operation}.

Discharge modes are employed to meet external control requests, supplying energy to the load or the grid. In the \textit{full disconnection mode}, the GiUPS is completely disconnected from the utility, and 100\% of the load power is supplied by the batteries (see S2 in Fig. \ref{UPS_GiUPS operation}). In the \textit{partial disconnection mode}, the input power from the grid is reduced according to external commands, with the remaining load power supplied by the batteries (see S3 in Fig. \ref{UPS_GiUPS operation}). For example, the batteries may provide 25\% of the total load power while the grid supplies the remainder.

When the battery state of charge (SOC) is below 100\% (e.g., 80\%) and an over-frequency event is detected, the GiUPS draws power from the grid to recharge the batteries. This operation is referred to as \textit{recharge mode}. The maximum recharge power is typically limited to 20--25\% of the GiUPS nominal capacity, constrained by battery recharge characteristics and the maximum input current.

The GiUPS can also operate as a bi-directional power converter to inject energy back into the grid (see S4 in Fig. \ref{UPS_GiUPS operation}). This \textit{energy export mode} is subject to local regulatory and grid requirements and allows the GiUPS to discharge batteries upstream. 

% The scope of grid interaction remains primarily focused on active power. Voltage regulation and power factor correction continue to be internal functions within the data center distribution system. This separation ensures that critical load protection is not compromised while enabling the UPS to contribute to grid stability.

% \vspace{-5mm}
\subsubsection{GiUPS Reactive Power Support and Voltage Ride-Through (VRT) Capability}
In addition to active power modulation, advanced GiUPS systems can provide reactive power injection and VRT capabilities, which are increasingly required by modern grid codes and interconnection standards~\cite{xie2025enhancing,azizi2026strengthening}.
% While traditional UPS systems focus primarily on maintaining voltage quality within the data center, GiUPS architectures extend this functionality to support upstream grid voltage stability~\cite{xie2025enhancing,azizi2026strengthening}.
Reactive power support is enabled by the bidirectional inverter stage, which can independently regulate active and reactive current components under a synchronous reference frame control strategy. GiUPS units typically provide reactive power capability in the range of 20--40\% of their rated kVA. For example, a 1~MW/1.1~MVA GiUPS can supply approximately $\pm$250--300~kVAr of dynamic reactive support.
% By adjusting the quadrature-axis current, the
GiUPS can inject or absorb reactive power without significantly affecting active power delivery to the critical load. This capability allows the DC to contribute to feeder voltage regulation, mitigate voltage fluctuations caused by rapid AI load variations (often inducing 5--10\% voltage swings within milliseconds), and improve local power factor at the point of common coupling (PCC) from typical values of 0.92--0.95 to above 0.99 \cite{VoltageRide-Through,xie2025enhancing}. 
% In weak grids with high renewable penetration, dynamic reactive power injection can enhance short-term voltage stability and reduce the likelihood of voltage collapse.

VRT capability ensures that the GiUPS remains connected and operational during short-duration voltage sags or swells~\cite{conto2025texas}. TUPS systems typically transfer to battery mode during severe disturbances, effectively isolating the load from the grid. In contrast, GiUPS systems are designed to comply with low-voltage ride-through (LVRT) and high-voltage ride-through (HVRT) requirements, similar to grid-scale inverter-based resources. Typical LVRT profiles require the inverter to remain connected down to 0.2~pu for 100--150~ms, and down to 0.5~pu for up to 500~ms. During a voltage sag, the GiUPS can increase reactive current injection proportionally to the voltage deviation, often delivering 2--3~pu reactive current within 2--5~ms, thereby supporting grid voltage recovery while maintaining uninterrupted supply to the critical load. 
% The inner current control loop ensures fast dynamic response, typically within a few milliseconds, while supervisory controls maintain battery state-of-charge (SoC) constraints and thermal limits.

The coordination between active power support, reactive power regulation, and VRT functionality requires a hierarchical control framework~\cite{zahedi2026best,li2016fault,cheng2011voltage}. In grid-following (GFL) mode, the GiUPS synchronizes to the grid using a phase-locked loop (PLL) and provides reactive current according to predefined droop characteristics (typically 3--5\% voltage droop) or grid operator commands. In grid-forming (GFM) mode, the inverter can regulate terminal voltage magnitude and frequency directly, enabling seamless transition to islanded operation if grid conditions deteriorate beyond acceptable thresholds. Protection logic ensures that grid support functions never compromise critical load reliability, maintaining load voltage within $\pm$1--2\% even during external disturbances.
% By incorporating reactive power control and ride-through capability, GiUPS systems evolve from passive backup devices into fully compliant inverter-based resources capable of supporting both frequency and voltage stability. This expanded functionality is particularly valuable in AI data centers, where high power density and rapid load transients interact with increasingly converter-dominated power systems.

\subsubsection{GiUPS Topologies}

% UPS systems can be implemented using several topologies, each with implications for efficiency, redundancy, and grid interaction capability~\cite{10255103}. Double-conversion online UPS architectures provide continuous conditioning and isolation but incur higher conversion losses. Delta conversion UPS architectures improve efficiency by allowing part of the power to flow directly from input to output while still providing regulation. Modular UPS architectures combine multiple smaller UPS modules to achieve the desired capacity and redundancy, allowing flexible scaling and selective participation in grid services.

For GiUPS applications, topologies that support bidirectional power flow and high-speed digital control are preferred. Double-conversion and delta conversion architectures with bidirectional converters can rapidly transition between load-following, grid-support, and islanded operation. Modular architectures are particularly attractive in AI DCs because they allow some modules to participate in grid services while others remain dedicated to critical load protection. This partitioning reduces operational risk and improves economic performance~\cite{Vertiv2025GridInteractiveUPS}. Emerging DC-coupled architectures further reduce conversion stages and improve round-trip efficiency. These architectures also enable direct coupling with on-site renewable generation or FCs, which enhances the ability of the AI DCs to operate as controllable grid resource.

\subsection{BESS Sizing in GiUPS Systems}

GiUPS BESS sizing must balance two objectives: maintaining backup reliability and delivering grid services such as FFR, primary regulation, and peak shaving \cite{ginzburg2025technical,rahman2026energy}. While backup needs establish the baseline energy capacity, grid-service participation adds important power-rating and cycling requirements, since services like frequency regulation are more power-intensive than energy-intensive. Economic sizing is therefore strongly influenced by cycle life, cumulative energy throughput, temperature, and average state of charge, because frequent shallow cycling can accelerate degradation \cite{pant2011introduction,kaiser2007optimized,zhang2019optimal,wang2026life,kamali2016life,omar2014lithium,majeau2011life,ding2019automotive}. Moderate oversizing can reduce effective DoD and extend battery lifetime, but excessive oversizing increases capital cost, footprint, and thermal management burden without proportional benefit. Accordingly, optimal sizing should jointly consider backup energy requirements, committed grid-service power capacity, and degradation-aware lifecycle cost modeling. For AI DCs, this process must also account for additional reserved capacity for short-term load smoothing, since this function helps mitigate upstream ramp rates and transformer stress but should not compromise guaranteed backup autonomy.

% ============================= BESS ==================================
% ===================== BESS =====================
% \subsection{Facility-Level Storage}

% \begin{itemize}
%     \item Role in damping large load fluctuations and supporting grid stability.
%     \item Control modes: Grid-forming vs. grid-following operation.
%     \item Use in peak shaving, frequency regulation, and black-start support.
%     \item Integration with distributed generation (DGs), renewables, and modular nuclear reactors.
%     \item Cost and sizing considerations for effective load smoothing.
% \end{itemize}\
% ===========================================================================
% \vspace{-3mm}
\section{BESS in AI DCs} \label{BESS}
BESS integrated with DCs represent a practical and scalable solution to address the challenges imposed by large and highly dynamic loads on the power system \cite{choukse2025power}. When appropriately designed and controlled, BESS can play a critical role in mitigating large load fluctuations, enhancing local power quality, and supporting overall grid stability \cite{NERC2025LargeLoads}. To establish a clear understanding of the underlying mechanisms and achievable benefits, this section reviews the fundamental BESS control paradigms, with emphasis on grid-forming and grid-following modes of operation. The discussion then extends to key grid-support functions enabled by AI DC–connected BESS, including co-generation, power smoothing and frequency regulation. Finally, the cost and sizing implications of BESS deployment are examined.

% ===========================================================================
% \vspace{-3mm}
\subsection{BESS Control Modes}
% BESS, enabled by power-electronic grid interfaces and flexible dispatch, are well suited to provide the dynamic active and reactive power support required for reliable grid operation while accommodating large-scale AI DC integration. In particular, GFM control is well matched to BESS applications because the DC-link provides an inherent energy buffer, allowing GFM functionality to be implemented primarily through software-based control strategies \cite{NERC2023GridFormingBESS}.

From an operational perspective, AI DCs can be configured for grid-tied or islanded operation, depending on the BESS inverter control mode and the facility-level coordination architecture. GFM control enables the BESS to establish the voltage and frequency reference required for islanded operation. In grid-tied mode, the BESS can charge or discharge to support peak shaving and increase renewable self-consumption, whereas in islanded mode it serves as the anchor resource to supply the AI DCs and potentially reduce reliance on conventional generation or support renewable co-generation \cite{NERC2023GridFormingBESS,SchneiderWP185BESS}.

\begin{table*}[!t]
\centering
\scriptsize
\caption{Summary of power-smoothing architectures for AI data centers: benefits and challenges \cite{QuantaTech2025AILoadProfiles}}
\label{tab:arch_benefits_challenges}
\renewcommand{\arraystretch}{1.15}
\setlength{\tabcolsep}{6pt}
\begin{tabular}{p{0.3\textwidth} p{0.25\textwidth} p{0.3\textwidth}}
\hline
\hline
\textbf{Architecture} & \textbf{Benefits} & \textbf{Challenges} \\
\hline
% \multirow{3}{*}{\textcolor{blue}{Layered ``Rack-Level'' + ``Grid-Level'' Solutions}} &
% \begin{itemize}\setlength{\itemsep}{1pt}\setlength{\parskip}{0pt}
% \item Load smoothing/shaping
% \item Demand flexibility
% \item Grid services
% \end{itemize}
% &
% \begin{itemize}\setlength{\itemsep}{1pt}\setlength{\parskip}{0pt}
% \item Higher cost
% \item Increased footprint
% \end{itemize}
% \\
% \hline
\multirow{4}{*}{Advanced GFM + GFL BESS Solution} &
\begin{itemize}\setlength{\itemsep}{1pt}\setlength{\parskip}{0pt}
\item Load smoothing/shaping
\item Demand flexibility
\item Grid services
\item Lower cost and footprint
\end{itemize}
&
\begin{itemize}\setlength{\itemsep}{1pt}\setlength{\parskip}{0pt}
\item Complex control strategy
\item Fast measurement and communication
\item Advanced plant-level control
\end{itemize}
\\
\hline
\multirow{4}{*}{Hybrid E-STATCOM + BESS Solution} &
\begin{itemize}\setlength{\itemsep}{1pt}\setlength{\parskip}{0pt}
\item Load smoothing and/or shaping
\item Demand flexibility
\item Grid services
\item Lower flicker
\end{itemize}
&
\begin{itemize}\setlength{\itemsep}{1pt}\setlength{\parskip}{0pt}
\item Higher cost
\item Increased footprint
\item Control interaction
\end{itemize}
\\
\hline
\multirow{3}{*}{Grid-Forming BESS} &
\begin{itemize}\setlength{\itemsep}{1pt}\setlength{\parskip}{0pt}
\item Load smoothing/shaping
\item Demand flexibility
\item Support grid services/stability
\end{itemize}
&
\begin{itemize}\setlength{\itemsep}{1pt}\setlength{\parskip}{0pt}
\item May require a line reactor
\item Higher flicker
\item Plant-level control
\end{itemize}
\\
\hline
\multirow{3}{*}{Supercapacitors (e.g., E-STATCOM)} &
\begin{itemize}\setlength{\itemsep}{1pt}\setlength{\parskip}{0pt}
\item High-speed response
\item Effective load smoothing/shaping
\item Reduced flicker
\end{itemize}
&
\begin{itemize}\setlength{\itemsep}{1pt}\setlength{\parskip}{0pt}
\item POI demand management
\item Not supporting grid services
\item Does not offer GFM support
\end{itemize}
\\
\hline
\hline
\end{tabular}
\vspace{-3mm}
\end{table*}

% ===========================================================================
% \vspace{-3mm}
\subsection{Grid Support Roles of BESS in AI DCs}
\subsubsection{On-Site Clean Power Integration}
% AI training workloads can trigger system resonances and severe voltage stress during abrupt ramps, especially in weak grids, underscoring the need for fast, power-electronics-based mitigation and coordinated control for disturbance-sensitive hyperscale AI DCs \cite{ko2025wide}. 

BESS play a critical role in enabling RESs integration at AI DCs. Given the energy-intensive nature of AI DCs, on-site renewable generation and off-site procurement mechanisms such as power purchase agreements (PPAs) are increasingly adopted to reduce operating costs and meet sustainability objectives. However, the inherent intermittency of renewable resources complicates reliable power supply. BESS mitigate these challenges by storing excess energy during periods of high generation and supplying it during low availability, while also providing backup power, and improved renewable utilization \cite{ahrabi2025ai}. AI DC–integrated BESS can be architected either as large, multi-megawatt GFM resources capable of replacing conventional backup generators, or as large-capacity batteries integrated within static UPS systems \cite{Eaton2025DataCentersGoodGridCitizen}. In this context, GiUPS systems, traditionally designed for short-duration ride-through, are increasingly complemented by dedicated BESS that are optimized for extended energy management and grid-interactive operation rather than transient backup alone \cite{SchneiderWP185BESS}. At the converter-control level, BESS can be configured for GFM, GFL, or hybrid GFM and GFL operation to balance fast dynamic response, voltage/frequency support, and grid-service capability, although with added control and sensing complexity.

\begin{figure}[!b]
    \centering
    \includegraphics[width=0.45\textwidth]{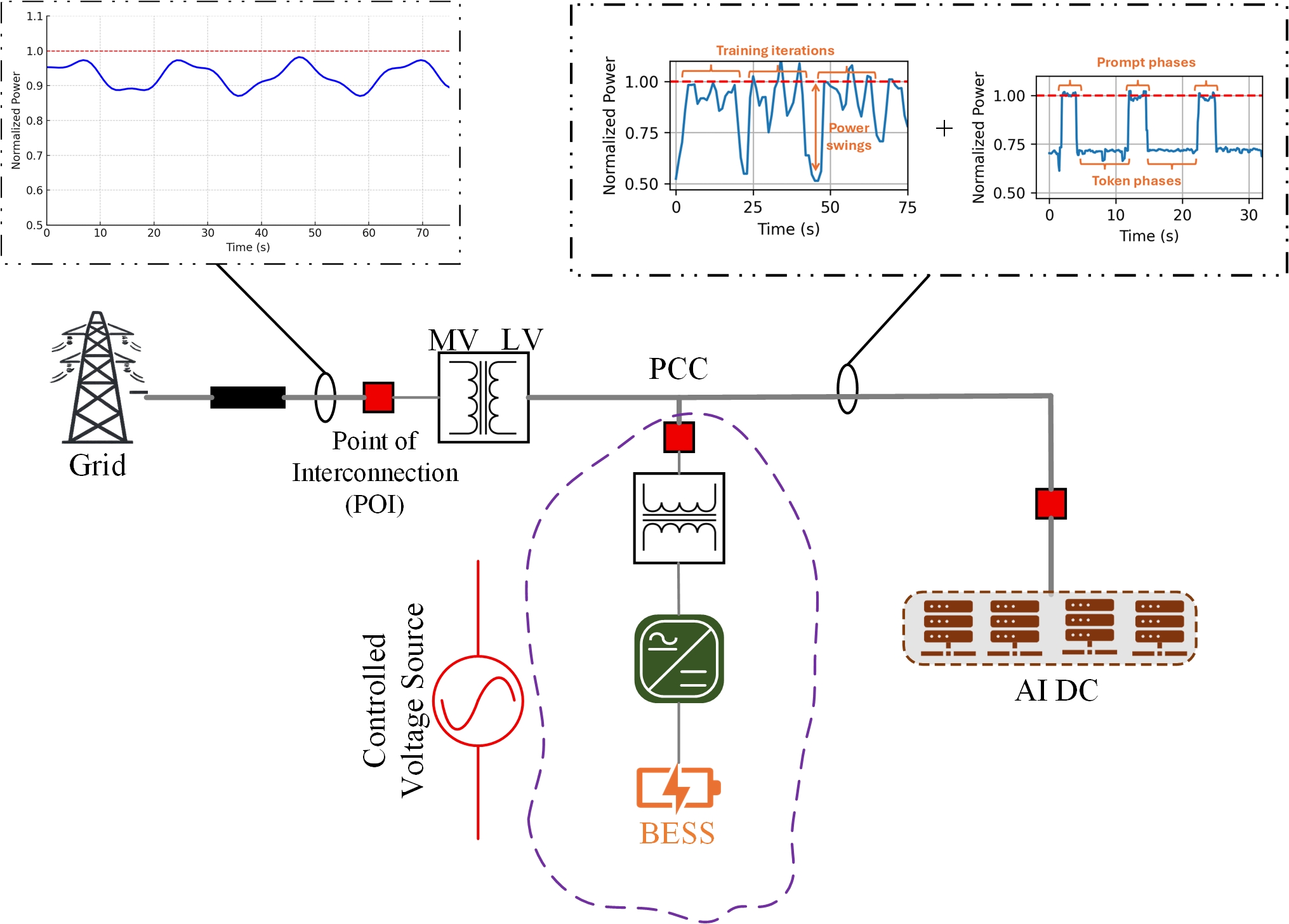}
    \caption{GFM BESS as a power smoothing unit \cite{QuantaTech2025AILoadProfiles}}
    \label{power smoothing}
    \vspace{-3mm}
\end{figure}

\subsubsection{Power Smoothing}
% Compared with generator-based solutions, which are constrained by emissions and slow start-up characteristics, BESS can respond rapidly and are therefore well suited for fast grid services such as frequency response and demand management \cite{paananen2021grid,Vertiv2025GridInteractiveUPS}. In particular, 

BESS can inject or absorb power at the facility interface to attenuate rapid AI DC demand variations and present a smoother power profile to the grid. A BESS co-located with an AI DC can provide grid-level power smoothing while supplying sufficient energy capacity to support sustained mitigation actions. This architecture improves operational reliability by enabling load shaping, demand flexibility, and extended ride-through for AI workloads. 

As shown in Fig. \ref{power smoothing}, the GFM BESS effectively smooths the power fluctuations of the AI DCs resulting from training and inference jobs. An alternative approach is a hybrid E-STATCOM and BESS configuration, which combines the fast, high-speed charge/discharge response of supercapacitors with the higher energy capacity of BESS to achieve effective smoothing and sustained load support. This hybrid architecture can enhance resilience by improving demand flexibility, reducing flicker, and enabling broader grid-service capability. The primary trade-off is increased cost and footprint compared with standalone solutions. In addition, standalone supercapacitor-based E-STATCOM solutions can mitigate high-frequency AI load fluctuations and improve power quality for sensitive IT equipment by reducing flicker, although they provide limited long-duration energy support \cite{QuantaTech2025AILoadProfiles}. Table \ref{tab:arch_benefits_challenges} compares the benefits and challenges of representative BESS-based and hybrid strategies for enabling grid support from AI DCs.

{\begin{figure}[!t]
\centering
\includegraphics[width=1\linewidth]{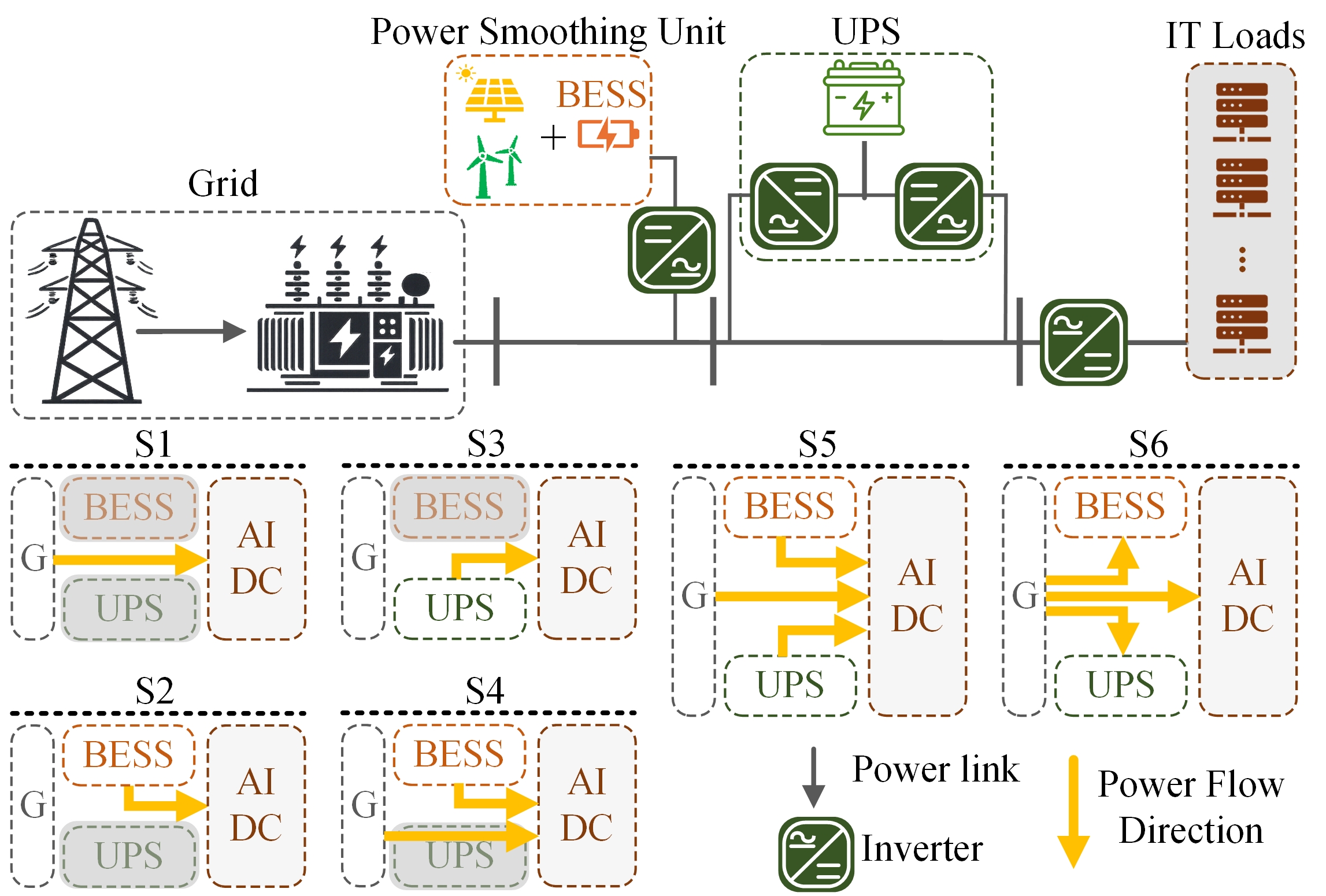}
\caption{UPS-Integrated BESS operation modes}
\label{Fig_IntegratedBESS_operation}
\vspace{-3mm}
\end{figure}}

% ===========================================================================
% \subsubsection{Grid Services Enabled by UPS-Integrated BESS}
\subsubsection{UPS-Integrated BESS}
UPS-integrated BESS are motivated by a simple premise: AI DCs already deploy UPS systems with fast power-electronic interfaces, and augmenting this mandatory infrastructure with BESS enables grid-interactive capability with limited additional integration complexity. In practice, grid-interactive deployments also show that battery technology selection depends on the targeted services and operating constraints. For example, UPS-integrated BESS have been used to provide FFR by leveraging the rapid controllability of UPS power converters \cite{Eaton2025DataCentersGoodGridCitizen,tao2025coordinated}. 

The UPS-integrated BESS, labeled as \emph{power smoothing units} in Fig. \ref{DC overview}, extend this concept by supporting long-duration outage operation and enabling participation in demand response programs, while the UPS maintains continuity of critical loads when neither the grid nor the BESS can fully supply demand. Although this architecture increases installation cost, it can significantly improve overall system reliability \cite{paananen2021grid}.

As summarized in Fig. \ref{Fig_IntegratedBESS_operation}, a GiUPS-integrated BESS can operate in (i) standard mode, (ii) discharge mode under full or partial disconnection, (iii) grid-services mode, and (iv) recharge mode for SoC recovery \cite{Vertiv2025GridInteractiveUPS}. This architecture can reduce the need for bidirectional power flow across the MV/LV distribution transformer by localizing fast power exchanges, allowing the transformer to operate primarily under its normal unidirectional loading conditions \cite{majeed2022impact}. These operating modes within the UPS--BESS subsystem enable long-duration BESS to enhance DC resilience through extended backup, reduce or eliminate reliance on diesel generation, participate in ancillary and reserve markets, manage demand charges and time-of-use tariffs, and increase RESs utilization \cite{SchneiderWP185BESS}.

\textit{S1) Normal operation:} The AI DC load is supplied entirely from the utility grid, and the system delivers 100\% of the input power to the AI DC.

\textit{S2) Full disconnection with sufficient BESS reserve:} The site is fully isolated from the utility, and upon command from an external controller, the BESS supplies 100\% of the AI DC load power. In this case, long-duration, large-scale BESS becomes essential; salt-cavern redox flow batteries are a promising due to their high safety, large storage capacity, stable temperature, and low cost \cite{pan2025salt}.  

During the transition from Scenario S1 to S2, the UPS initially supports the load because its response time is on the order of milliseconds, while the BESS typically requires a few seconds to assume full load supply. Once engaged, the BESS can sustain operation for extended durations (e.g., on the order of 1--4 hours, depending on sizing and operating conditions) \cite{SchneiderWP185BESS}.

\textit{S3) Full disconnection with insufficient BESS reserve:} If the outage duration is long and the BESS does not have sufficient available capacity to supply the full AI DC load, the UPS assumes responsibility for supporting the critical load.

\textit{S4) Partial disconnection:} The external controller reduces the grid input power, and the remaining portion of the AI DC demand is supplied by the BESS. This operating mode enables the BESS to support grid frequency regulation and to present a smoother power profile at the grid interface.

\textit{S5) Coordinated UPS--BESS power smoothing:} The grid, BESS, and UPS jointly supply the AI DC load, with the UPS helping the BESS by providing fast support during critical transient conditions to assist power smoothing and frequency response. This case can be interpreted as an extension of S4 in which the UPS contributes rapid, short-duration buffering when required.

\textit{S6) Recharge mode:} When the SOC drops below full charge, the utility initiates battery recharging (e.g., following an over-frequency detection). Under this condition, the UPS is prioritized for recharging first, after which the BESS can be recharged from the grid or from on-site renewable generation.

\subsection{BESS Cost and Sizing Considerations}
The cost of BESS per kilowatt-hour has declined dramatically, dropping from approximately \$7,500 in 1991 to \$133 in 2024, with projections indicating a further reduction to around \$80 by 2030 \cite{SchneiderWP185BESS}. This downward cost trajectory is expected to continue with further scale-up in manufacturing capacity and advances in both manufacturing and battery recycling processes.
% Thus, multiple cost-effective grid-scale BESSs can be implemented in future AI DCs.
% Reference \cite{8217257} shows that a 560 kWh Lithium-ion BESS, along with a 960 kWh lead-acid BESS, can provide FFR and backup power generation for TDCs. This study can be further extended to AI DCs as well, in which the cooperation between UPS and BESS is also considered.

In addition to capital cost, effective sizing must account for operational objectives. Co-generation, power smoothing, and assisting GiUPS each require independent studies to determine the optimal BESS size. Sizing also affects the number of BESS units required for AI DCs. The sizing study will provide precise information on whether a single hyperscale BESS is more optimal than multiple grid-scale BESS units. Recent studies have applied advanced methods to determine the optimal BESS capacity while ensuring the required capacity and satisfying dispatch constraints \cite{mamun2016multi,ye2026grid,yu2025reliability}. Reference \cite{8217257} shows that a 560 kWh Lithium-ion BESS, along with a 960 kWh lead-acid BESS, can provide FFR and backup power generation for TDCs. This study can be further extended to AI DCs as well, in which the cooperation between UPS and BESS is also considered.

\section{Fuel Cells (FCs) in AI DCs} \label{FC}
FCs convert chemical energy directly into electrical energy through electrochemical oxidation reactions. Because this conversion is not combustion-based, FCs can achieve electrical efficiencies exceeding 60\%, which is significantly higher than many conventional generators~\cite{fan2021recent}. FC systems offer high operational reliability due to their modular design and minimal use of mechanical components, resulting in reduced maintenance requirements and high availability. When supplied with hydrogen, FCs produce no 
carbon dioxide or conventional air pollutants during operation~\cite{manzo2025fuel}.

{\begin{figure}[t]
\centering
\includegraphics[width=\linewidth]{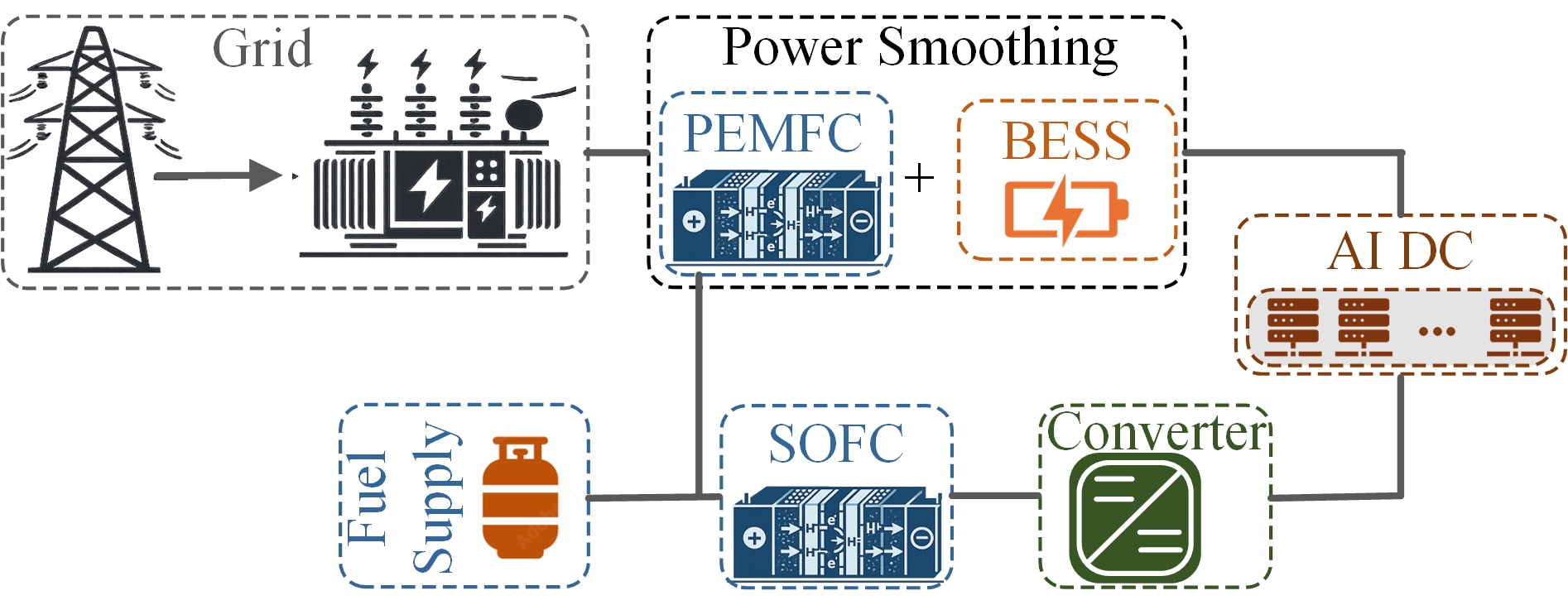}
\caption{FC integration with AI DC.}
\label{Fig_FCinAIDC}
\vspace{-3mm}
\end{figure}}

Among the several established FC chemistries, Solid Oxide Fuel Cells (SOFCs), in context of AI DCs, are often highlighted for their high steady-state efficiency and suitability for long-duration on-site power generation~\cite{nikiforakis2025understanding}. Polymer Electrolyte Membrane Fuel Cells (PEMFCs) exhibit faster electrochemical response and greater compatibility with power-electronic interfacing, making them promising candidates for coordinated power smoothing~\cite{kong2025grid}. As illustrated in in Fig. \ref{Fig_FCinAIDC}, both SOFC and PEMFC systems rely on a shared upstream fuel supply while serving distinct operational roles: SOFCs assist with base-load power to the AI DC, whereas PEMFCs, co-located with BESS, operate at faster timescales to mitigate short-term power fluctuations. With a consistent and reliable fuel supply, this architecture enables hierarchical coordination between the fuel infrastructure, on-site generation, and the utility grid.

% From an economic perspective, the adoption of FCs depends on total cost of ownership, including stack and balance-of-plant capital costs, hydrogen storage and delivery infrastructure, and the delivered cost of hydrogen.

%However, PEMFC stacks are subject to ramp-rate limitations to avoid fuel starvation and efficiency degradation, which constrains their ability to directly follow rapid load transients.%
\vspace{-2mm}
\subsection{Role of FC in AI DCs}
Commercial deployments, such as Bloom Energy’s SOFC systems, despite their relatively high capital cost, have demonstrated the feasibility of FC installations capable of supplying primary 
or supplemental power to TDCs. These systems are typically configured to provide continuous or long-duration power, serving as low-emission and more energy-efficient alternatives to diesel generators~\cite{perez2025enabling}. 
In emerging AI DCs, whose power profiles are characterized by large, rapid, and highly synchronized load variations, FC-based technologies must be integrated within a broader energy architecture and 
operated as slow-following or baseload resources in order to preserve efficiency and lifetime. Accordingly, two complementary roles are considered:

\subsubsection{FC On-Site Generation}
SOFC-based systems are primarily suited for on-site generation in AI DC environments due to their high steady-state electrical efficiency, fuel flexibility, and scalability to megawatt-class power levels \cite{nikiforakis2025understanding}. When supplied by pipeline natural gas or on-site hydrogen storage, SOFC systems can provide sustained power delivery over extended durations, making them alternatives to diesel generators for backup and resilience applications. Unlike battery-based systems, whose backup duration is limited by stored energy capacity, FC-based backup duration is primarily constrained by fuel availability, enabling outage coverage ranging from hours to multiple days depending on infrastructure configuration. However, due to their high operating temperatures and slow thermal dynamics, SOFC systems are not designed for rapid power modulation and are therefore best operated as baseload or slow-following generators within a coordinated energy management framework.

\subsubsection{FC-Assisted Power Smoothing}
PEMFCs operate at significantly lower temperatures than SOFCs and exhibit faster electrochemical response, enabling limited power modulation relative to high-temperature FC technologies. Experimental studies at the tens-of-kilowatt scale have demonstrated that PEMFC systems can achieve sub-second power ramping under carefully controlled conditions \cite{nikiforow2018power}. However, these studies also show that stack power extraction must be constrained by maximum allowable ramp rates to prevent fuel starvation, voltage collapse, and accelerated degradation. Repeated exposure to aggressive load transients has been associated with catalyst deterioration, membrane stress, and performance loss. Consequently, while PEMFCs may contribute to low-frequency power balancing service provision, they are not well suited to serve as primary resources for high-frequency load smoothing in AI DCs \cite{li2025ai}.

In practical deployments, PEMFC operation must, therefore, be coordinated with on-site BESSs and GiUPS systems. In such hierarchical architectures, BESSs and UPS systems regulate fast load transients and maintain AI DC-bus stability on millisecond-to-second timescales, while PEMFCs are dispatched through filtered power references that limit ramp rates and confine operation to degradation-aware regimes. Continuous monitoring of PEMFC operating conditions is essential both for degradation analysis and for coordinated control with fast-acting storage systems. When integrated in this manner, PEMFCs can complement BESS-dominated load smoothing strategies by supplying average power and extending overall system endurance without being subjected to damaging high-frequency load variations \cite{kong2025grid}. 

% \vspace{-3mm}
\subsection{Sizing Considerations}

Scaling FC technologies to AI DC capacities introduces constraints related to physical aggregation, hydrogen logistics, and balance-of-plant complexity.

GFM control strategies for PEMFC systems have primarily been validated at the tens-of-kilowatt scale. Scaling such architectures to the multi-megawatt level requires parallel stack aggregation, coordinated hydrogen distribution manifolds, auxiliary subsystem synchronization, and high-capacity DC/DC and DC/AC conversion stages \cite{zheng2025techno}. 

SOFC systems exhibit a distinct scaling trajectory. Hybridized configurations that integrate bottoming cycles for thermal energy recovery have demonstrated stable operation at multi-megawatt and hundred-megawatt scales~\cite{nikiforakis2025understanding}. In these installations, performance gains arise primarily from thermodynamic integration. As capacity grows, system complexity (increases) toward thermal management and mechanical integration.

Accordingly, sizing decisions for AI DC deployments should not be based solely on nominal power ratings, but on the interaction between electrochemical constraints, infrastructure requirements, and load temporal characteristics. PEMFC capacity, if deployed, should be bounded by hydrogen logistics and ramp-rate considerations, while SOFC capacity may be aligned with the steady baseload fraction of demand. 

% normalized cost being nevertheless determined over $0.10/kWh
%Due to the information reported in Sections 5.1 Power Generation, 5.5 Findings, documented efficiencies correlated to applications exclusively providing electrical power output, ranging from 5 kWe to 627 MW 

% ======================== TES ============================
% \vspace{-3mm}
\section{Thermal Energy Storage (TES) in AI DCs} \label{TE}
TES refers to technologies that store energy in the form of heat or cold by raising or lowering the temperature of a storage medium relative to ambient conditions \cite{alva2018overview}.  Unlike electrochemical and mechanical energy storage, TES does not directly support the electrical power, instead it reshapes the energy demand by decoupling thermal energy production from its consumption. This makes TES relevant in systems where thermal load are tightly coupled with performance and electrical demand, such as DCs \cite{enescu2020thermal}.

% \subsection{Role of Thermal Energy Storage in AI Data Centers}

In AI DCs, cooling constitutes  30-40\% of total facility energy consumption \cite{omrani2025ai} due to the high power density of accelerator-based computing. Nearly all electrical energy consumed by GPUs and associated IT equipment is ultimately converted into low-grade heat, with studies estimating that around 90 \% of input electrical power manifests as waste heat within the DC environment \cite{omrani2025ai}. Cooling demand, as a result, scales with computational load. While training DC power exhibits high-frequency power swings, thermal load evolves on slower timescales. This creates favorable conditions for the use of thermal energy storage to store cooling capacity and supply it during periods of high electrical demand or grid stress.

% \vspace{-3mm}
\subsection{Characteristics and Timescale of TES Operation}
TES exhibits operational characteristics that differ fundamentally from electrical ESSs. Due to thermal inertia and heat transfer dynamics, these systems respond on slower timescales, typically ranging from several minutes to hours. As a result, TES is not suited for fast grid services such as frequency regulation or power quality support \cite{guelpa2019thermal}. 

Instead, the storage units are well matched to applications involving load shifting, peak demand reduction, and ramp-rate mitigation of cooling-related electrical loads. Thermal standby losses and insulation constraints further limit the suitability of TES for long-term or seasonal storage in most data center applications. Consequently, TES is most effective when operated on diurnal timescales, complementing faster electrical storage technologies within a hierarchical energy storage framework \cite{enescu2020thermal}. 

% \vspace{-3mm}
\subsection{TES Technologies and Integration Architectures}
\subsubsection{Sensible TES}

Sensible TES stores cooling capacity through temperature changes in a storage medium, most commonly water. Chilled water storage tanks are a mature and widely deployed form of the technology due to their simplicity, low cost, and ease of integration with conventional chilled-water systems \cite{sarbu2018comprehensive}.  However, because energy is stored through temperature differential rather than phase change, sensible TES exhibits comparatively lower volumetric energy density than alternative technologies. As a result, achieving equivalent storage capacity requires significantly larger physical volume and installation area. 

Despite this footprint requirement, sensible water-based systems remain the least expensive option per unit of stored thermal energy and demonstrate predictable thermal behavior with minimal operational complexity \cite{guelpa2019thermal}. For AI DC campuses where land availability permits, sensible TES provides a cost-effective and technically robust solution for short-term load shifting and peak chiller reduction.

\subsubsection{Latent TES (PCM-Based)}

Latent TES exploits the phase change of materials to store thermal energy at nearly constant temperature. Compared to sensible storage, phase change materials (PCMs) provide higher volumetric energy density, allowing equivalent cooling capacity to be achieved with substantially reduced installation volume. This characteristic is particularly advantageous in space-constrained AI DC facilities \cite{sarbu2018comprehensive}.

The most widespread latent storage medium in cooling networks is ice, owing to its high latent heat and relatively low material cost. Ice-based TES has been successfully deployed in large-scale district cooling systems worldwide, demonstrating technical feasibility at substantial capacity \cite{sarbu2018comprehensive}. However, latent systems introduce additional engineering complexity, particularly in heat exchanger integration, phase stability, and long-term cycling reliability. Although more compact than sensible storage, latent TES generally entails higher capital cost per unit of stored energy. Consequently, while ice-based TES offers clear volumetric advantages for cooling applications, its deployment in AI DC environments must balance footprint reduction against increased system complexity and economic uncertainty.

\subsubsection{Thermochemichal TES}
Thermochemical TES stores energy through reversible chemical or sorption processes, allowing thermal energy to be stored as chemical potential rather than as temperature change. Such systems can achieve high energy density and the lowest standby losses, making them attractive for long-duration storage \cite{sarbu2018comprehensive}. 

Despite these advantages, thermochemical TES systems are generally complex, expensive, and less mature than sensible or latent storage technologies. Their application in DCs is currently limited to research and demonstration projects, though they represent a potential future option as cooling demands continue to increase.

% \vspace{-3mm}
\subsection{TES for RESs Integration}
TES can support the integration of RESs in AI DCs by enabling power-to-cooling strategies. During periods of surplus RESs generation or low electricity prices, chillers can be operated to charge TES systems, storing cooling capacity for later use \cite{luerssen2020life}.  This approach improves the utilization of RESs and reduces reliance on grid power during peak or high-emission periods.

By shifting cooling-related electrical demand in time, TES enhances the flexibility of AI DCs as large grid-connected loads and supports broader decarbonization objectives.

% ===========================================================================

% \subsection{Intermediate-Level Storage}
% \begin{itemize}
%     \item UPS (Uninterruptible Power Supply) Systems:
%     \begin{itemize}
%         \item Operation modes and communication with the grid.
%         \item Unidirectional vs. bidirectional power flow capabilities.
%         \item Grid-interactive UPS: supporting ancillary services and islanded operation.
%         \item Cost and reliability comparison between UPS and on-site BESS.
%     \end{itemize}
%     \item Thermal Storage Units:
%     \begin{itemize}
%         \item Integration with server cooling.
%         \item Coordination with electrical storage for workload management.
%     \end{itemize}
% \end{itemize}

% \vspace{-5mm}
\section{Rack- and Server-Level ESSs roles in AI DCs} \label{rack}
\begin{figure}[t]
\centering
\includegraphics[width=0.8\linewidth]{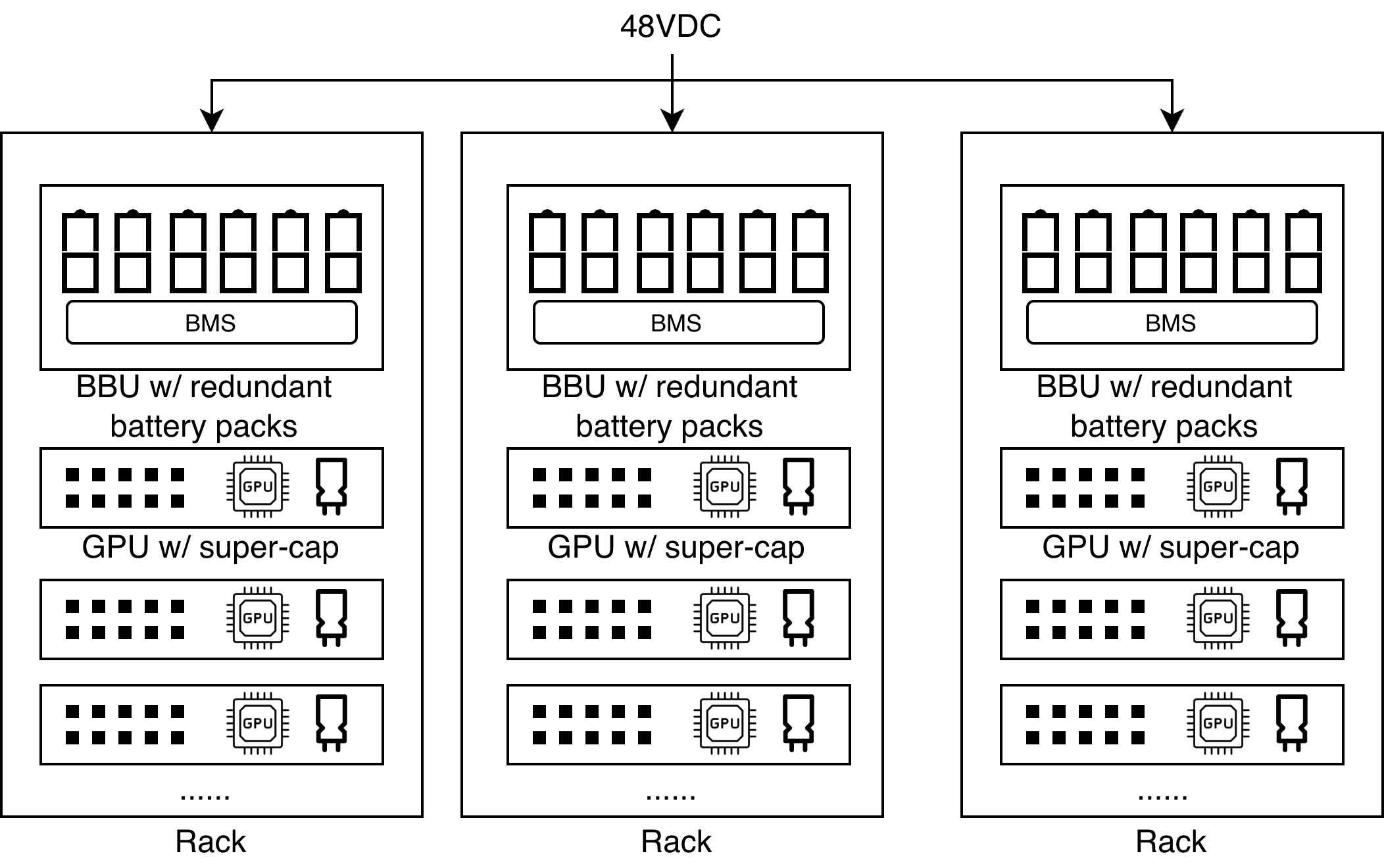}
\caption{Rack-level BBUs and server-level capacitors}
\label{Fig_rack}
\vspace{-3mm}
\end{figure}

Rack- and server-level ESSs exist to handle two constraints that centralized UPS/BESS alone cannot satisfy: (i) \emph{local buffering} (GPU-/server-scale transients that require local energy buffering), and (ii) \emph{fault domain} (limiting the blast radius of power disturbances and energy storage failures). In modern DC-architecture AI facilities, upstream layers (SST/MVDC front-ends and facility UPS/BESS) primarily set facility envelopes and ride-through at longer timescales, while downstream layers move buffering closer to the load to smoothen fast dynamics and localize contingencies under faults\cite{800vdc_nvidia,iec_mvdc_overview}. This section focuses on rack-level BBUs as distributed ride-through and backup resources, and on server/GPU-level capacitive storage as power profile smoothening for the highest-frequency behaviors\cite{ocp_orv3_bbu_module_spec_Rev1.4,ocp_orv3_bbu_shelf_spec_Rev1.1,nvidia_gb300_power}.
% \begin{table*}[t]
% \centering
% \tiny
% \caption{High-level comparison of rack-level BBUs and centralized UPS systems.}
% % \begin{tabularx}{\columnwidth}{p{2.2cm} X X}
% \begin{tabular}{ccc}
% \hline
% \textbf{Aspect} & \textbf{Rack BBUs} & \textbf{Central UPS}\\
% \hline
% Primary purpose 
% & Rack ride-through; local fault containment \cite{ocp_orv3_bbu_shelf_spec_Rev1.1}
% & Facility ride-through; centralized management \cite{eaton_microsoft_energy_aware_ups} \\

% Timescales 
% & Seconds--minutes (backup, recovery) \cite{adi_ocp_orv3_bbu_reference} 
% & Minutes+ (backup) and grid services where allowed \cite{microsoft_dublin_ups_grid} \\

% Sub-second smoothing 
% & Not ideal; accelerates battery aging \cite{aging_aware_liion_review,li2025ai} 
% & Not ideal; GPU transients are localized rail events \cite{li2025ai} \\

% Coordination burden 
% & High; fleet SoC, recharge scheduling, priority handling \cite{micro20_priority_charging} 
% & Lower; centralized components \\

% Efficiency 
% & Avoids some double conversion \cite{safari25efficiency} 
% & Often involves AC--DC--AC in classic designs \\
% \hline
% % \end{tabularx}
% \end{tabular}
% \label{tab:bbu_vs_ups}
% \vspace{-3mm}
% \end{table*}

\begin{table}[b]
\centering
\scriptsize
\caption{High-level comparison of rack-level BBUs and centralized UPS systems.}
\begin{tabular}{p{2.2cm} p{2.6cm} p{2.6cm}}
\hline
\hline
\textbf{Aspect} & \textbf{Rack BBUs} & \textbf{Central UPS} \\
\hline
Primary purpose 
& Rack ride-through; local fault containment \cite{ocp_orv3_bbu_shelf_spec_Rev1.1}
& Facility ride-through; centralized management \cite{eaton_microsoft_energy_aware_ups} \\

Timescales 
& Seconds--minutes (backup, recovery) \cite{adi_ocp_orv3_bbu_reference}
& Minutes+ (backup) and grid services where allowed \cite{microsoft_dublin_ups_grid} \\

Sub-second smoothing 
& Not ideal; accelerates battery aging \cite{aging_aware_liion_review,li2025ai}
& Not ideal; GPU transients are localized rail events \cite{li2025ai} \\

Coordination burden 
& High; fleet SoC, recharge scheduling, priority handling \cite{micro20_priority_charging}
& Lower; centralized components \\

Efficiency 
& Avoids some double conversion \cite{safari25efficiency}
& Often involves AC--DC--AC in classic designs \\
\hline
\hline
\end{tabular}
\label{tab:bbu_vs_ups}
\vspace{-3mm}
\end{table}

% \vspace{-3mm}
\subsection{Rack-Level ESS: Role, Benefits, and Constraints}
The rack represents a natural operational and electrical boundary for AI infrastructure, with dense DC-architecture racks alone reaching MW-scale power demands \cite{800vdc_nvidia}. Deploying energy storage at the rack level offers several advantages. First, the storage is electrically close to the load, enabling local buffering before propagating upstream to the facility distribution system. Second, failures in battery, converters, or wiring can be contained to individual racks, limiting the blast radius of faults. Third, storage capacity budgets and ride-through priorities can be configured on a per-rack basis, allowing workload-specific policies that reflect criticality differences across the DC.

Despite these benefits, rack-level deployment introduces operational challenges. Coordination becomes necessary, as many independent storage units must be managed for state-of-charge tracking, recharge scheduling, and availability monitoring. Safety and compliance requirements increase, since colocation of lithium-ion energy storage with IT equipment raises fire and thermal-runaway mitigation concerns. Finally, maintenance overhead grows substantially, as consistent monitoring, battery replacement, and preventive maintenance must be performed across racks. Fig. \ref{Fig_rack} shows the implementation of rack-level BBUs and server-level capacitors in the IT equipment.

\subsection{Battery Backup Units (BBUs) as Distributed UPS}
In OCP Open Rack designs, BBUs are integrated into the rack as distributed DC ride-through modules that supply the rack DC bus during upstream AC disturbances \cite{ocp_orv3_bbu_module_spec_Rev1.4}. In Open Rack V3 (ORV3) architecture, a BBU shelf typically hosts multiple BBU modules with \(5{+}1\) redundancy and supports both \emph{charge mode} and \emph{discharge mode} operations with monitoring of SoC/SoH and maintenance tests \cite{ocp_orv3_bbu_shelf_spec_Rev1.1}. Public reference designs summarize the intended operating point (e.g., per-module backup power on the order of kW for minutes, and lower-power charging over hours), reflecting a design goal of short ride-through and fast recovery rather than continuous power profile smoothening \cite{adi_ocp_orv3_bbu_reference}. Meta reports ORV3 BBU shelves designed for minutes of backup, and notes that paired shelves can be used for higher rack power configurations \cite{meta_orv3_grand_teton_ocp}.

\subsubsection{BBU vs centralized UPS}
BBUs and centralized UPS systems can both respond quickly in principle to supply the gap between grid power disturbance or outage and longer-term energy storage systems coming online (e.g., gensets / BESS), but they differ in \emph{where} energy is buffered and \emph{what} the buffer is optimized for. Centralized UPS concentrates energy and power conversion at room/facility scale, simplifying management and enabling grid-interactive use cases when permitted \cite{eaton_microsoft_energy_aware_ups,microsoft_dublin_ups_grid}. Rack BBUs push storage behind fewer conversion stages in DC racks and reduce some double-conversion penalties compared to traditional centralized UPS architectures \cite{safari25efficiency}. Practically, BBUs are best viewed as a distributed energy availability layer for the rack (seconds--minutes), not as the primary solution for sub-second smoothing which is handled more effectively by server/GPU capacitive buffers. Table \ref{tab:bbu_vs_ups} summarizes the high level comparison of rack-level BBUs and centralized UPS systems.

\subsubsection{BBUs as power absorbers}
BBU shelves inherently include charging control; in normal operation they remain in charge mode and can modulate charging current/power within design limits \cite{ocp_orv3_bbu_module_spec_Rev1.4}. At hyperscale, uncontrolled simultaneous recharge can create step increases in aggregate load that trip upstream protection. Production case studies show that coordinated and priority-aware charging of distributed batteries can dramatically reduce recharge power (reported reductions up to \(\sim\)80\%) while meeting recharge constraints \cite{micro20_priority_charging}.  

However, using BBUs as a frequent, high-rate smoothing solution can accelerate degradation depending on cycling throughput, C-rate, SoC range, and temperature \cite{aging_aware_liion_review,crawford_grid_duty_cycles}. Practically, server/GPU capacitors and software solution are better suited to handle high-frequency power fluctuation, and BBUs are better suited for ride-through, controlled recovery, and low-frequency shaping with budgeted cycling.

\subsubsection{Safety and deployment considerations}
Rack BBUs introduce Li-ion safety considerations near IT equipment. OCP BBU specifications explicitly reference safety and propagation constraints and relevant certification/testing practices \cite{ocp_orv3_bbu_module_spec_Rev1.4,ul9540a}. This strongly motivates designs that (i) enforce conservative SoC windows and thermal monitoring, and (ii) reduce unnecessary cycling.  

If the design intent is ``UPS-like'' ride-through for all computing devices, deployment of BBU on every rack is the straightforward approach used in OCP-style architectures \cite{ocp_orv3_bbu_module_spec_Rev1.4,ocp_orv3_bbu_shelf_spec_Rev1.1,meta_orv3_grand_teton_ocp}. Selective deployment (e.g., BBUs only on the highest-power or highest-priority racks) can reduce capex and coordination overhead, but it changes the availability model: unprotected racks become dependent on workload-level fault tolerance, or upstream ESS (BESS/gensets) without transient protection.

% \vspace{-5mm}
\subsection{Server- and GPU-Level Storage}
\subsubsection{Capacitors instead of batteries}
The highest-frequency components of AI load power dynamics are at server-level: rapid power consumption swings on the GPUs. These events are handled by local decoupling and bulk capacitance \cite{choukse2025power}. Buffering at this layer prevents microsecond-to-millisecond disturbances from propagating to the power sharing unit (PSU) and rack bus, reducing upstream stress and allowing higher utilization at the rack power infrastructure \cite{li2025ai}.

Compared with batteries, supercapacitors tolerate extremely frequent charge--discharge cycles with minimal degradation, making them attractive for sub-second smoothing \cite{ieee_spectrum_supercaps_ai,eaton_supercapacitor_ai}. Industry offerings and proposals include compact supercapacitor banks intended to suppress short spikes that would otherwise require overbuilding upstream power infrastructure. The main constraint is energy density and physical volume: supercapacitors can buffer transients and short bursts, but they cannot provide minutes of ride-through and therefore complement (not replace) rack BBUs and other upstream ESSs \cite{eaton_supercapacitor_ai}.

\subsubsection{Firmware/software shaping as energy ``storage''}
A key trend is co-design of electrical buffering with firmware-level power shaping. Vendors explicitly target smoother power draw by controlling ramp rates and limiting transient excursions. For example, NVIDIA describes firmware-controlled ramp-up behavior and a ``power burner'' mode to manage ramp-down and stabilize facility-level power dynamics during AI job transitions \cite{HowNewGB300NVL72FeaturesProvideSteadyPowerforAI}. At the platform level, software interfaces such as NVIDIA’s NVML expose enforced GPU power limits and real-time power telemetry, enabling operators and higher-level controllers to constrain GPU power draw and coordinate device-level consumption with rack- or facility-level power envelopes
 \cite{nvidia_nvml}. It is framed as an industry direction to approach power stabilization as a cross-stack problem that couples hardware energy storage and software control to prevent large excursions and meet grid/facility ramp constraints \cite{choukse2025power}.

\subsection{Sizing Considerations for Rack- and Server-Level ESSs}
Sizing at the rack and server layers is driven by different objectives than grid-scale or UPS-level storage. Rather than energy capacity for extended outages or grid-service participation, the dominant constraints are rack power consumption, backup duration target, and redundancy requirements.

The fundamental BBU shelf sizing requirement is that the shelf must supply the full rack power demand for the target ride-through duration under a single module failure. For AI racks, where power demands can exceed 100~kW and reach MW-scale in dense configurations~\cite{800vdc_nvidia}, this places a substantially larger energy requirement on the shelf than in traditional deployments. In ORV3 architectures, the 5+1 redundancy model requires each active module to be sized for its proportional share of the full rack load plus margin~\cite{ocp_orv3_bbu_shelf_spec_Rev1.1}. For target backup duration, OCP reference designs reflect a goal of minutes of ride-through ~\cite{adi_ocp_orv3_bbu_reference,meta_orv3_grand_teton_ocp}. The specific duration that rack-level storage supports should be designed in sync with upstream ESSs, namely BESS or gensets, since its task is to bridge outage onset with extended ESS being online. Extending this window increases energy capacity but consumes already limited rack space, as well as amplifying thermal risks for Li-ion cells colocated with IT equipment~\cite{aging_aware_liion_review,ocp_orv3_bbu_module_spec_Rev1.4}. 

Supercapacitors at the server level are sized to absorb the transient energy of rapid GPU power swings~\cite{li2025ai}. For sub-second transients, while power swings could be large, the energy required is limited, which makes supercapacitors viable at this layer despite their low energy density~\cite{choukse2025power,ieee_spectrum_supercaps_ai}. The binding constraint is therefore capacity with regard to the GPU power swing rather than cycle life, since supercapacitors tolerate very high cycle counts with minimal degradation~\cite{eaton_supercapacitor_ai, HowNewGB300NVL72FeaturesProvideSteadyPowerforAI}.

% \vspace{-5mm}
\subsection{Coordination Across Rack and Server Layers}
Coordination should be explicit about objectives and timescales: GPU/server controllers shape fast dynamics under performance constraints; rack BBU controllers manage SoC availability and recovery without creating recharge spikes at facility level; facility controllers enforce facility envelopes and power-quality limits \cite{micro20_priority_charging,harmonics_review_2025}. OCP rack ecosystems already expose the necessary hooks (telemetry, shelf controllers, parallel shelf operation) to implement multi-layer coordination in practice \cite{ocp_orv3_bbu_module_spec_Rev1.4,ocp_orv3_bbu_shelf_spec_Rev1.1}. Under MVDC architectures, server/GPU capacitive buffering can reduce required BBU power bandwidth and cycling, but does not eliminate the seconds--minutes energy role of BBUs for ride-through and controlled recovery \cite{iec_mvdc_overview}.

% % ===========================================================================
% \textcolor{blue}{\textbf{[Revision — BEGIN]}}
% \input{BESS/BESS_Rack_Level}
% \textcolor{blue}{\textbf{[Revision — END]}}
% % ===========================================================================

% \section{Converter and Control Topologies}
% \begin{itemize}
%     \item AC/DC, DC/DC, and AC/DC/AC conversion stages across hierarchical layers.
%     \item Converter Topologies:
%     \begin{itemize}
%         \item Power sharing, bidirectional flow, and power density optimization.
%         \item Grid-forming vs. grid-following converters in microgrid-oriented designs.
%         \item Solid-State Transformers (SSTs) as potential alternatives for modularity and bidirectional control.
%     \end{itemize}
%     \item Control Strategies:
%     \begin{itemize}
%         \item Power sharing, load balancing, and hierarchical coordination.
%         \item Multi-timescale control (millisecond-level GPU buffers to minute-level BESS operations).
%         \item Data-driven and AI-based control for predictive load management.
%     \end{itemize}
% \end{itemize}

{\begin{figure}[t]
\centering
\includegraphics[width=0.95\linewidth]{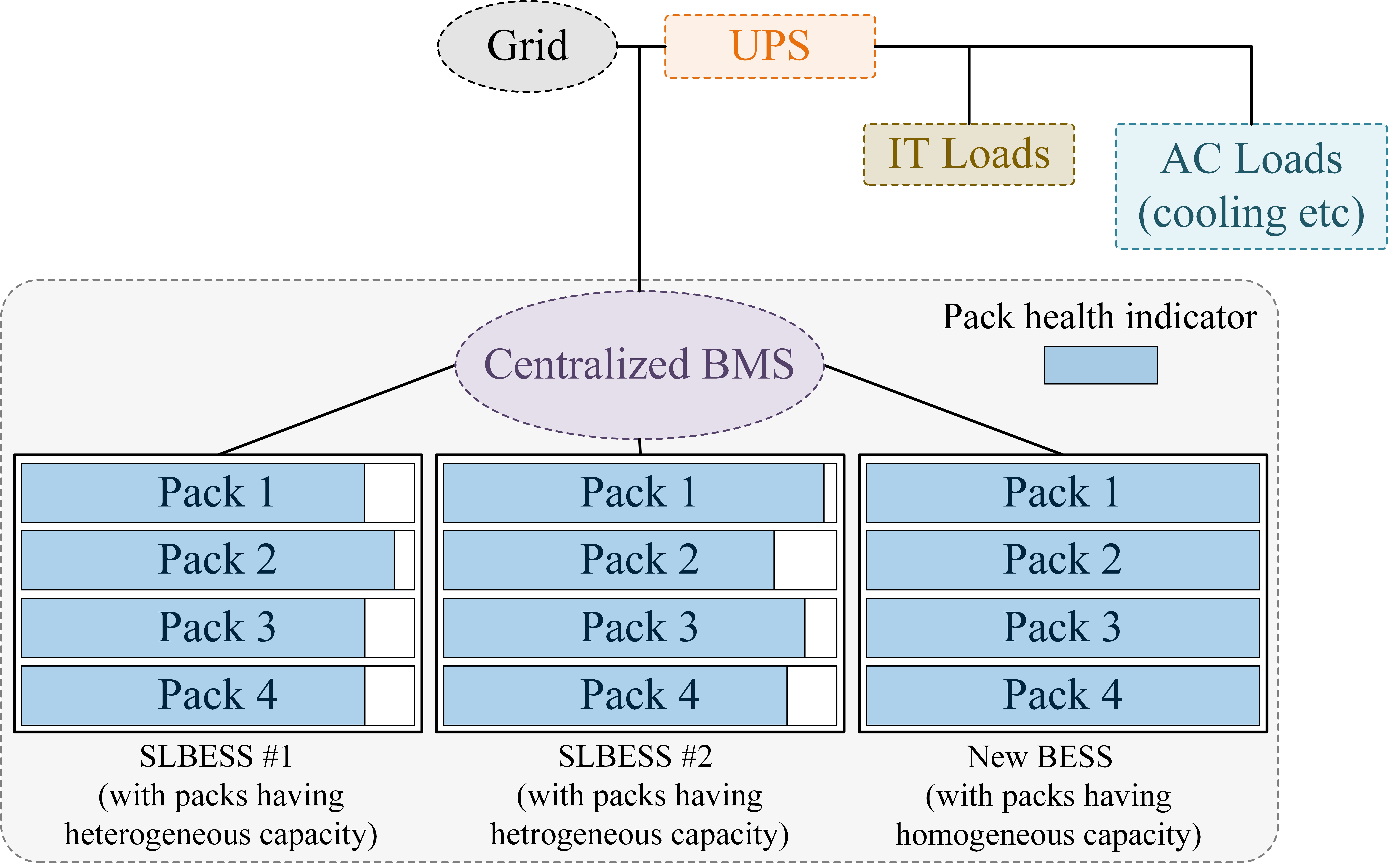}
\caption{SLBESS with heterogeneous composition for backup generation of UPS for AI DC load.}
\label{Fig_SLBESS}
\vspace{-3mm}
\end{figure}}

\vspace{-3mm}
\section{Second-Life Battery Energy Storage Systems (SLBESS) in AI DCs} \label{SLBESS}
% \begin{itemize}
%     \item Potential for replacing gensets and extending UPS backup capacity.
%     \item Degradation modeling and characterization of second-life cells.
%     \item Challenges in control, state-of-health estimation, and safety.
%     \item Cost–reliability trade-offs for integrating second-life storage at different hierarchy levels.
%     \item Standards and policies for reuse in data centers.
% \end{itemize}

Having established the functional roles of ESSs across architectural layers, an important practical question concerns cost scalability, motivating the use of second-life batteries (SLBs).

Electric vehicle (EV) batteries are typically retired when they reach around 80\% of their initial capacity. Retired batteries can either be sent to recycling facilities or repurposed for secondary applications, giving them a second life. One of the most common second-life applications is stationary BESS for the electric power grid~\cite{Hassan2023SecondLifeBatteries}.  Many U.S. and Canada based startups have raised funding to re-purpose used batteries. 
For example, B2U Storage Solutions~\cite{B2UStorage}, Element Energy~\cite{ElementEnergy}, Moment Energy~\cite{MomentEnergy}, 
Smartville~\cite{Smartville}, and Redwood Energy~\cite{Redwood2025AIStorage} have each demonstrated the technical and economic viability of SLBs as behind-the-meter (BTM) and front-of-the-meter (FTM) BESS. Beyond technical feasibility and cost savings, the reuse of SLBs confers substantial environmental advantages by decreasing lifecycle emissions and supporting a circular economy~\cite{Bobba2018SecondLifeLCA}. While the application of SLBESS for grid support is well documented, their deployment for AI DC loads is an emerging and promising opportunity, as highlighted in a recent project by Crusoe Energy, where the company installed used batteries to power AI DC load~\cite{CrusoeRedwood2025Microgrid}.
% AI DCs must comply with building codes, electrical regulations, and safety standards such as 
% NFPA~70E~\cite{NFPA70E}, 
% NFPA~101~\cite{NFPA101}, 
% and ANSI/TIA-942-C~\cite{TIA942}.
SLBESS installations must also adhere to battery repurposing standards such as UL 1974~\cite{UL1974}. Therefore, regulatory compliance is mandatory for building AI DC infrastructure and battery reuse.

%\textcolor{blue}{Retired EV batteries with varying cell chemistry can be utilized as SLBESS; however, LFP batteries are particularly useful for second-life applications over other types (e.g., NMC or NiMH) due to their enhanced life cycle and lower recycling value ~\cite{Bach2025FairMarketValue}.}

% \vspace{-3mm}
\subsection{SLBESS Assisting AI DC On-Site Generation}
For AI DC load, SLBESS can serve as BTM generation backup source, provided that the battery packs are rigorously tested and qualified via health metrics. This is a GFL mode of operation where hybrid configurations, combining used and new battery packs, are technically feasible as shown in Fig. \ref{Fig_SLBESS}. The battery management system (BMS) balances the varying capacities of different packs. Additionally, a single SLB pack exhibits heterogeneity across cells and modules, resulting in variable capacities, voltages, and current limits. E.g. a single degraded cell may constrain the performance of an entire module or pack. To solve this, passive and active balancing schemes have been explored to mitigate these challenges. Advanced control architectures have been proposed to actively balance heterogeneous cell capacities in SLBs~\cite{Wang2023LifeExtendedControl}.

% \vspace{-3mm}
\subsection{Hybrid  GiUPS configuration for AI DCs using SLBESS }

UPS systems in AI DCs provide instantaneous backup for critical IT loads. This necessitates energy storage solutions capable of sustained discharge durations exceeding 10 hours. SLBESS are generally not suitable for extended durations, and operate within a narrow SoC window to avert deep discharge. This is because the over-discharge accelerates cell-level degradation and may push cells to the non-linear ``knee point'' region in the degradation curve. Thus, unless carefully derated and managed, SLBESS may offer limited suitability for long-duration UPS applications.

In AI DC, using SLBESS for short-duration backups while avoiding rapid degradation is technically feasible. Optimal dispatch strategies must be implemented to achieve that. Traditionally, the fundamental approach of using BESS involves energy arbitrage, where the BESS charging occurs when electricity prices are low, and discharge occurs when the electricity prices are high. This economic objective is balanced against the physical aging of the battery through a degradation cost component, shown in Eq. \ref{eq:optimal_dispatch_arbitrage_time}.

\begin{equation}
\label{eq:optimal_dispatch_arbitrage_time}
\max \left(
C_{dis,t}\, P_{dis,t}
-
C_{ch,t}\, P_{ch,t}
\right)
-
C_{deg}\!\left(\text{cycles}\!\left(P_{dis,t}, P_{ch,t}\right)\right)
\end{equation}

\noindent subject to:
\begin{itemize}
    \item Battery Constraints that include limits on the SoC and maximum charge/discharge rates to ensure operational safety.
    \item Grid Constraints that includes power exchange limits and local grid regulations.
\end{itemize}

For ever time step, $t$ in hours, $C_{dis}$ and $C_{ch}$ are the market price (\$/kWh) of electricity during the discharge and charge phase. $P_{dis}$ and $P_{ch}$ are the power output (kW) discharged from and charged to the SLBESS. $C_{deg}$ is the degradation cost coefficient, representing the cost of battery wear per cycle. The cycles of BESS can be calculated as a function of aggregated energy throughput $f(P_{dis}$,$P_{ch})$. The sum of $P_{ch}$ and $P_{dis}$ represents the total power throughput, which serves as a proxy for the stress placed on the SLBESS. To find $C_{deg}$ in \$/cycle, the data from an accelerated life testing of SLB cells of corresponding chemistry can be used.

For AI DCs, the primary objective of energy storage optimal dispatch shifts from pure arbitrage to supply guarantee and reliability. In a hybrid GiUPS configuration, the SLBESS must prioritize the load over economic gains. The optimality equation can be modified to include a reliability term, which serves as a significant penalty for unserved load, ensuring the AI DC remains operational. Therefore, Eq. \ref{eq:optimal_dispatch_arbitrage_time} can be modified as Eq. \ref{eq:obj5}

% \begin{equation}
% \label{eq:obj5}
% \max \Bigg[
% \big(C_{\mathrm{dis},t} P_{\mathrm{dis},t} - C_{\mathrm{ch},t} P_{\mathrm{ch},t}\big)
% - C_{\mathrm{deg}}\big(P_{\mathrm{ch},t} + P_{\mathrm{dis},t}\big)
% - C_{\mathrm{reliability}}\,L_{\mathrm{unserved},t}
% \Bigg]
% \end{equation}

\begin{equation}
\label{eq:obj5}
\max \Bigg(
\begin{aligned}
& C_{\mathrm{dis},t} P_{\mathrm{dis},t}
- C_{\mathrm{ch},t} P_{\mathrm{ch},t} \\
& - C_{\mathrm{deg}}\!\left(P_{\mathrm{ch},t}+P_{\mathrm{dis},t}\right)
- C_{\mathrm{reliability}} L_{\mathrm{unserved},t}
\end{aligned}
\Bigg)
\end{equation}

\begin{figure}[t]
\centering
\includegraphics[width=0.9\linewidth]{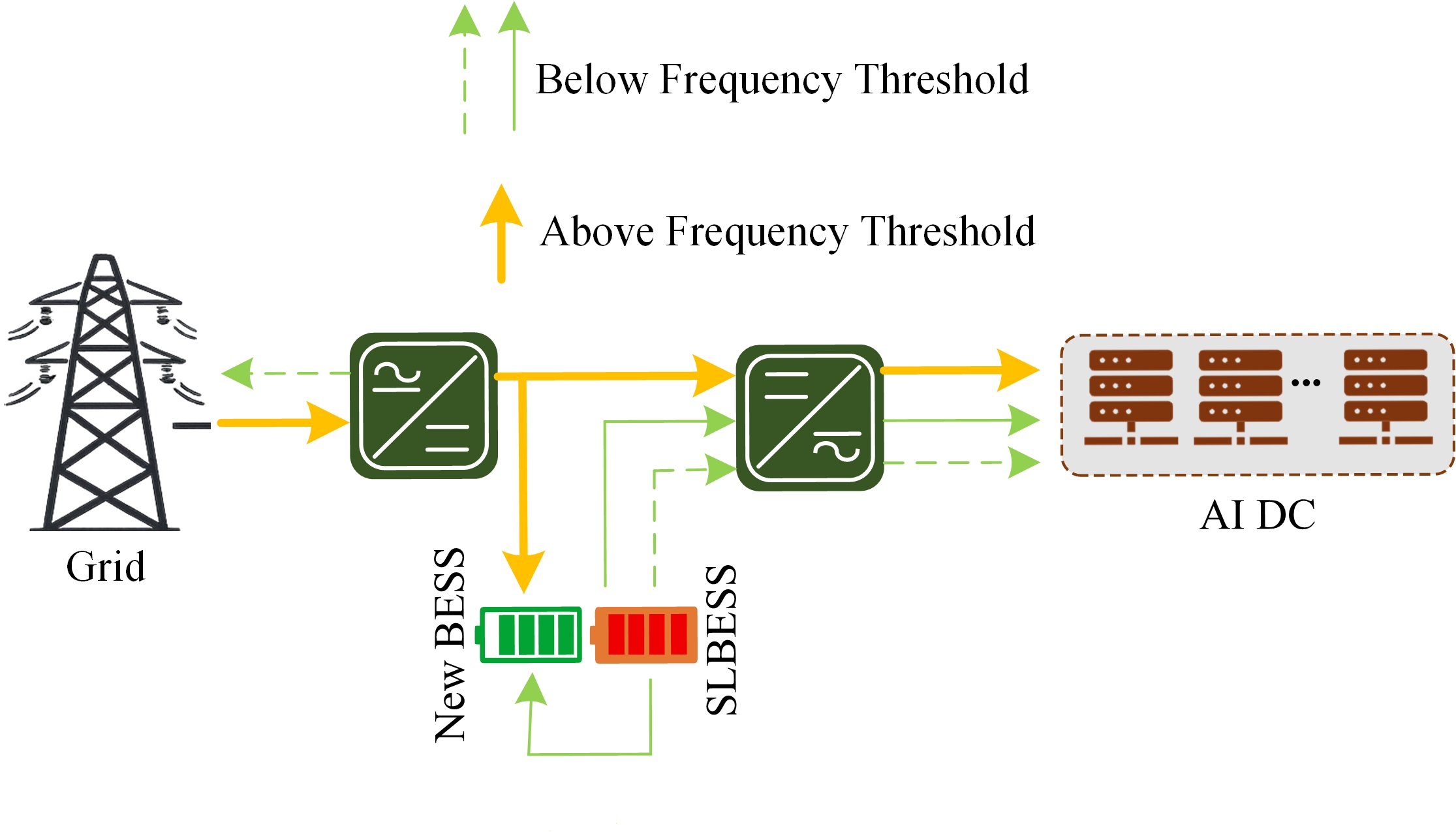}
\caption{FFR by utilizing new BESS and SLBESS in GiUPS architecture}
\label{FFR}
\vspace{-5mm}
\end{figure}

\noindent $C_{reliability}$ is a high-value penalty factor associated with the loss of power to the AI DC (interpreting the "supply guarantee" requirement from the sources) and $L_{unserved}$ is the portion of the critical AI DC load not met by the grid or the SLBESS while subjected to the BESS and UPS constraints.

On the grid side, GiUPS can be a source of FFR. This is possible with bidirectional rectifiers that can respond to changing grid frequency. FFR tends to use the BESS frequently within a narrow SoC window. SLBESS alone is not suitable for this sort of grid service, leading to its rapid degradation \cite{White2020RepurposedSLBFR}. However, if used alongside a new BESS, the UPS controller can be programmed to leverage the SLBESS for charging the new BESS when its SoC falls below a certain threshold. If optimized, this can reduce the energy import from the grid, saving the cost of energy. Fig. \ref{FFR} shows the possible scenarios, which are listed below.
\\
\begin{itemize}
    \vspace{-3mm}
    \item[(i)] In case of grid frequency above the threshold, the UPS controller charges the new BESS and SLBESS, providing stabilization of the frequency.
    \item[(ii)] In case of grid frequency below the threshold,
    \\
    a) The UPS controller can discharge new BESS and provide power back to the grid while powering the IT load with SLBESS. 
    \\
    b)The UPS controller can meet a part of the IT load using the new BESS, and use SLBESS to charge the new BESS and stop any import/export to the grid.
\end{itemize}

% \subsection{Second-life-Battery  Degradation, State Estimation, and Remaining Useful Life (RUL) Prediction during operation}

% Lithium-ion batteries undergo both calendar degradation and cyclic aging~\cite{Kostenko2024SecondLifeAging}. Calendar degradation is time-dependent and occurs regardless of battery usage, while cyclic degradation results from charge-discharge cycles. Degradation pathways, including solid electrolyte interphase (SEI) growth, lithium plating, and particle fracture~\cite{Hassan2023SecondLifeBatteries} lead to capacity fade and increased internal resistance. To avoid the SLBs from entering the non-linear region of the degradation curve, a degradation-aware dispatch strategy needs to be devised.

% Estimating the remaining useful life (RUL) of SLBs is essential for safe deployment. Approaches range from data-driven techniques~\cite{Cui2025BMS2}, neural network-based models, to entropy-based algorithms~\cite{Strugnell2024EntropyBattery}. Uniquely, aging curves of SLBs are shaped by first-life usage patterns, making every battery distinct. European Union (EU) led Initiative  Battery Passport~\cite{EU2023BatteryPassport} tracks lifecycle data, empowering refurbishers to select candidates for reuse.

% \vspace{-3mm}
\subsection{Grid-Scale SLBESS for AI DCs}
The US national grid has a long queue for interconnection requests. This leads to limited generation and difficulties in accommodating new AI DC load. In \cite{norris2025rethinking}, the authors concluded that 76 GW of new load can be integrated without significant upgrades by curtailing 0.25\% of new flexible load per year. This amounts to a curtailment of 22 hours/year. FTM grid-scale SLBESS can achieve this by shaving peaks of AI DC load, decreasing the stress on the grid that is caused by erratic peaks of AI DC load.
It can also acts as standby generation and partake in resource adequacy. Furthermore, as hyperscalers such as Meta, Microsoft, and  Apple Inc.
increasingly engage in energy market operations \cite{Roy2025BigTechEnergy}, grid-scale FTM SLBESS can be utilized to manage surplus power from renewable generation plants under PPAs enabling these hyperscalers to hedge against increasing energy prices.

By implementing the optimal dispatch strategies previously discussed, these systems can store excess renewable energy during periods of high production or low market cost and sell it back to the wholesale market when prices are high. This strategy is particularly effective when using lithium-ion batteries, such as lithium iron phosphate (LFP), which have been proven to be a viable and high-capacity option for DC loads.

However, to successfully integrate this FTM SLBESS approach into hyperscale infrastructure, several factors must be carefully balanced:

\begin{itemize}
    \item \textbf{Reliability vs. Profit:} While selling energy to the wholesale market provides economic benefits, the dispatch logic must prioritize supply guarantees and the reliability of the AI DC load over arbitrage profits.
    
    \item \textbf{Degradation Management:} Because these systems utilize second-life battery cells, dispatch strategies must be carefully managed, and aging of the SLBESS must be monitored more rigorously than conventional BESS to prevent accelerated degradation during market participation.
    
    \item \textbf{System Optimization:} Hyperscalers can leverage real-time SLBESS status to optimize hybrid BESS/UPS system operation, enabling effective peak shaving and grid interaction while maintaining sufficient power reserves for critical AI DC operations.
    \vspace{-3mm}
\end{itemize}

% \vspace{-3mm}
\subsection{Cost and Sizing Aspects of SLBESS for AI DCs}
\subsubsection{Csot} SLBESS is typically more cost-effective than new BESS. However, cost is influenced by supply, location, and market conditions. The cost of stationary storage declined from \$169/kWh in 2020 to about \$108/kWh in 2025, amounting to 36\% reduction worldwide~\cite{BNEF2025BatteryCost}. This decline trajectory may be a hindrance to SLBESS usage in the future. Moreover, 
refurbishing, operation, and maintenance can constitute a significant portion of CapEx. Market analysis suggests LFP batteries are more economically advantageous for second-life use compared to nickel-rich chemistries, primarily due to the NCX battery's higher recycling value posing a demand competition between SLB repurposers and recyclers~\cite{Bach2025SecondLifeValue}. Since reliability can be a huge factor in AI DCs, the hyperscalers may prioritize new BESS over SLBESS. However,  the retirement of EV batteries is expected to coincide with rising DC energy demand and the policies, such as the \textit{One, Big, Beautiful Bill Act} (OBBBA) bill, introduce \textit{investment tax credits} (ITC) for battery storage sourced within the country, enhancing the attractiveness of domestically available SLBs for AI DC deployment~\cite{IRSOBBB}.

\subsubsection{Sizing}Sizing SLBESS for AI DCs presents a distinct engineering challenge. SLBs, repurposed from EV applications, exhibit reduced usable capacity, increased internal resistance, and greater heterogeneity in state-of-health (SoH) compared to new cells \cite{casals2019second,martinez2018battery}. To mitigate accelerated degradation and ensure operational reliability, SLBs are typically operated within narrower SoC windows and under derated conditions \cite{neubauer2011ability}. Unlike new batteries that can safely utilize a broad operational SoC range (e.g., 10–90\%), second-life systems are often constrained to tighter windows (e.g., 30–70\%) to limit stress and extend remaining useful life. This restriction directly reduces the effective usable energy fraction, thereby requiring oversizing in both energy (kWh) and power (kW) capacity relative to the predicted load demand of AI DCs.
% AI DCs impose stringent reliability requirements. These facilities must maintain high availability under both normal and contingency conditions, typically requiring N+1 or 2N redundancy configurations. Consequently,
SLBESS must be sized not only for operational objectives such as FFR, peak shaving, and short-duration backup, but also to satisfy redundancy and fault-tolerance criteria. The SoH variability across SLBs introduces capacity dispersion and imbalance, which must be explicitly accounted for in system-level sizing methodologies \cite{casals2019second}.
% Statistical characterization of SoH distributions and degradation trajectories is therefore necessary to determine the minimum installed capacity required to guarantee a specified effective capacity with high confidence over the planning horizon.

Thermal and cycling considerations further complicate the sizing process. For SLBs with limited remaining useful life (RUL), constraining DoD and per-cycle energy throughput becomes essential to mitigate capacity fade and impedance growth. However, these operational safeguards further reduce usable capacity and necessitate additional oversizing. Moreover, increased internal resistance in aged lithium-ion cells leads to reduced power capability, particularly at lower SoC levels \cite{vetter2005ageing}. This reinforces the requirement for power oversizing to ensure that transient load spikes can be supported without violating voltage, thermal, or safety constraints.
% Therefore, sizing SLBESS for AI DCs must integrate degradation-aware operational limits, SoH uncertainty, redundancy requirements, and dynamic load characteristics within a unified techno-economic framework.

% \vspace{-3mm}
% \subsection{Standards and Policies for Battery Reuse in AI DCs}

% AI DCs must comply with building codes, electrical regulations, and safety standards such as 
% NFPA~70E~\cite{NFPA70E}, 
% NFPA~101~\cite{NFPA101}, 
% and ANSI/TIA-942-C~\cite{TIA942}. SLBESS installations must also adhere to battery repurposing standards such as UL 1974~\cite{UL1974}. Therefore, Regulatory compliance is mandatory for building AI DC infrastructure and battery reuse.

% \section{Comparative Assessment and Design Trade-offs}
% \begin{itemize}
%     \item Cost, complexity, reliability, and efficiency trade-offs.
%     \item Performance metrics: response time, lifetime, degradation rate, and energy density.
%     \item Design optimization between power density and energy capacity.
%     \item Economic feasibility of integrating fuel cells, nuclear microreactors, and renewables.
% \end{itemize}

\vspace{-3mm}
\section{Challenges and Future Directions} \label{conclusion_challenges}

This section synthesizes the challenges and future research directions arising from the multi-layer ESS integration reviewed in this paper. We first address current challenges for both AI DC operation and ESS integration, then discuss future technical directions. Table~\ref{tab:ess_comparison} provides a consolidated cross-technology comparison.

\begin{table*}[t]
\centering
\scriptsize
\caption{Comparative Summary of Energy Storage Solutions for AI DCs}
\label{tab:ess_comparison}
\setlength{\tabcolsep}{3pt}
\renewcommand{\arraystretch}{0.98}
\resizebox{\textwidth}{!}{%
\begin{tabular}{p{1.9cm}p{1.6cm}p{1.7cm}p{2.5cm}p{2.7cm}p{1.8cm}p{2.1cm}p{2.3cm}}
\toprule
\textbf{Technology} & \textbf{Response Time} & \textbf{Application Level} & \textbf{Key Advantages} & \textbf{Key Challenges} & \textbf{Relative Cost} & \textbf{AI DC Role} & \textbf{Grid Interaction} \\
\midrule
Chip-level capacitors / HBM buffers & Sub-ms to ms & Chip & Eliminates GPU power spikes; protects sensitive hardware & Limited energy capacity; thermal management & Low per-unit; high aggregate & GPU power smoothing & None (local) \\
\midrule
Rack BBU / supercapacitors & ms to seconds & Rack/Server & Fast transient response; modular deployment & Limited energy; degradation under high-cycle loads & Moderate & Server backup; rack-level buffering & None (local) \\
\midrule
TUPS & $<20$ ms & Facility & Proven reliability; load isolation & Passive operation; idle batteries; no grid services & Moderate--High & Load ride-through & Minimal \\
\midrule
GiUPS & 0.5--10 s (FFR) & Facility & Bidirectional operation; active grid support; FFR/VRT capability & Complex control; regulatory approval; battery stress & High & FFR, VRT, peak shaving, frequency regulation & Active (FFR, VRT, peak shaving) \\
\midrule
BTM BESS & Minutes--hours & DC-scale & Demand charge reduction; RES integration; backup support & High CapEx; large footprint; sizing complexity & High & Load smoothing; backup; RES support & Moderate (via PCC) \\
\midrule
FTM BESS & Minutes--hours & Grid-scale & Large capacity; ancillary services; reserve support & Permitting; grid interconnection; high CapEx & Very High & Grid reserve; emergency supply & Strong \\
\midrule
Fuel Cell (FC) & Seconds--minutes & Facility & Low-carbon operation (H$_2$); high energy density; diesel replacement & Fuel supply chain; cost; H$_2$ infrastructure & High & Clean backup generation; co-generation & Moderate \\
\midrule
TES & Minutes--hours & Facility & Cooling load shifting; RES utilization; no electrochemical degradation & Integration complexity; auxiliary power; space requirement & Moderate & Cooling efficiency; peak load reduction & Indirect \\
\midrule
SLBESS & Seconds--hours & DC-scale / facility & Lower CapEx; circular economy; scalable deployment & SoH uncertainty; cell heterogeneity; safety standards & Low--Moderate & Cost-effective BTM/FTM backup & Moderate \\
\bottomrule
\end{tabular}%
}
\end{table*}

\subsection{Current Challenges for AI Data Centers}

\subsubsection{AI DC Load Characteristics and Grid Impact}
As discussed in the Introduction and Section~\ref{power architecture}, AI DC load profiles differ fundamentally from traditional loads. The sub-second power variability of training workloads causes harmonic distortion, voltage flicker, and sub-frequency mechanical oscillations in synchronous generators \cite{CharacteristicsandRisksofEmergingLargeLoads,10255103}. These impacts are amplified when multiple GPU clusters train synchronously. At the transmission level, the sudden switch of large AI DCs to backup power, as demonstrated by the 1,500 MW event in Virginia \cite{VirginiaDataCenter}, reveals that current grid planning frameworks are insufficient for accommodating these highly responsive large loads. The primary challenge is the combination of high power magnitude, rapid and poorly predictable load transitions, and the absence of coordinated grid-response protocols between DC operators and utility grid operators.

\subsubsection{Simulation and Experimental Validation}
To validate the optimal operation of the energy management system (EMS) of an AI DC, it is necessary to model its different components, including the ESSs. To the best of the authors' knowledge, there is no software specifically developed for simulating EMSs of AI DCs. While MATLAB/Simulink or GT-SUITE can integrate mechanical and electrical components, the multi-physics nature of AI DCs, requiring simultaneous transient analysis for power electronics and thermal units alongside steady-state analysis for grid-scale systems, is not well supported by existing tools \cite{cao2025transforming,li2025phythesis,colangelo2025ai}. Available real-time simulator devices are similarly not suitable for real-time experimental analysis of AI DCs, since they are primarily developed for electrical circuits rather than combined electro-thermal-mechanical systems. This gap limits the ability to validate hierarchical ESS coordination strategies before physical deployment.

\subsubsection{GPU Scheduling and Energy-Aware Compute}
Current GPU-based AI workloads are predominantly executed at the highest supported core frequency, prioritizing peak performance at the expense of energy efficiency. Unlike CPUs, where DVFS is a mature and widely adopted mechanism, energy-aware frequency regulation for GPUs remains at an early stage \cite{chung2026joules,wu2026kareus}. Existing GPU scheduling policies largely optimize for throughput and latency without explicitly accounting for the energy consumption characteristics of heterogeneous deep learning tasks \cite{chung2024reducing,chung2025ml}. The resulting lack of controllability over GPU power profiles directly constrains the effectiveness of chip-level and rack-level ESSs in power smoothing applications, since these systems must cope with power profiles that are neither observable nor controllable from the ESS perspective.

\subsubsection{Distribution Transformers and Bidirectional Power Flow}
Traditional distribution transformers are designed for unidirectional power flow. AI DCs equipped with large-scale ESSs and advanced power electronics can dynamically shift between high consumption and active power injection for grid services. Under bidirectional power flow, conventional transformers experience accelerated thermal aging due to loading patterns that deviate from standard design assumptions \cite{majeed2022impact}. Voltage regulation mechanisms and tap settings optimized for passive load behavior may become ineffective or unstable during rapid transitions between load and injection states \cite{FrotscherRaveTeNyenhuisUpadhyay2021RPF}. These limitations underscore the need for enhanced transformer models, real-time monitoring, and power-electronic-aware grid interfaces at AI DC interconnection points.

\subsubsection{Load Forecasting for ESS Sizing and Dispatch}
Traditional grid loads have predictable trajectories that can be forecast using historical data. AI DC load profiles are fundamentally different: training job submissions, checkpoint saves, and idle periods create highly stochastic power demand that cannot be reliably modeled with conventional forecasting techniques. As a result, ESS sizing methodologies based on historical load patterns may under- or over-estimate required capacity. Rack-level BESSs and chip-level batteries designed for power smoothing require accurate sub-second power consumption data, while grid-scale BESSs require accurate load profiles at the facility level for effective grid support. Advanced machine learning-based load forecasting methods \cite{lu2026dynamic,mughees2025short,jiang2025hyperload,mo2025learning} and high-resolution measurement infrastructure \cite{chalamala2025data} are essential prerequisites for reliable ESS design and operation in AI DCs.

\subsubsection{Advanced ESS Degradation Modeling and Lifetime Prediction}
Due to high power variability in AI DCs, more frequent and irregular charging and discharging cycles for ESSs are inevitable. This requires precise degradation analysis to maximize lifetime and optimize ESS operation. As discussed in Section~\ref{rack}, ESSs at the chip or rack level experience high-frequency cycling under varying conditions to smooth the power profile. Online monitoring can support better lifetime prediction and proactive maintenance. For lithium-ion batteries, cyclic aging is a key concern at both the small scale (BBU, chip-level) and large scale (BESS). SLBESSs are particularly vulnerable to aging effects and require more rigorous monitoring, especially for larger units where cell-level heterogeneity amplifies uncertainty in state-of-health estimation \cite{Kostenko2024SecondLifeAging,Hassan2023SecondLifeBatteries}.

\subsubsection{Hierarchical ESS Coordination}
Multi-layer ESS implementation in AI DCs requires a centralized EMS to monitor and coordinate operating modes across all ESS layers \cite{li2025ups_datacenter}. The on-site grid-scale BESS must operate coherently with the GiUPS to enable accurate power sharing, RES utilization, and emergency grid support. The GiUPS battery size and switching time are fundamentally coupled to the size of the on-site BESS. This relationship requires combined reliability and cost analyses. The work \cite{ko2025mitigation} demonstrates a novel control framework utilizing supercapacitors and grid-scale BESS for load smoothing, illustrating the potential of hierarchical coordination. TES systems must also operate coherently with the BESS and FC. BBUs and chip-level ESSs must coordinate for multi-layer power smoothing. These interdependencies necessitate a centralized control and management system with appropriate data acquisition infrastructure spanning all layers.

\subsubsection{Long-Duration Energy Storage (LDES)}
On-site large-scale BESSs and GiUPS systems require long-duration energy storage (LDES) capability to compensate renewable intermittency or to sustain AI DC operation during extended utility outages \cite{11230052}. Current LDES technologies are not yet mature for cost-effective deployment at the scale required by hyperscale AI DCs, and further investigation is needed into chemistries, system integration, and control strategies suitable for these applications.

\subsubsection{Optimal Sizing and Cost Considerations}
There is limited public research on the optimal sizing and cost analysis of multi-layer ESSs in AI DCs. For instance, it is unclear whether a grid-scale BESS is more cost-effective than a GiUPS system with high power-density batteries, or whether large-scale supercapacitors at the rack level are preferable to a combination of BBUs and supercapacitors in load-smoothing scenarios. Answering these questions requires multi-objective techno-economic frameworks that integrate backup energy constraints, committed grid-service capacity, degradation-aware lifecycle cost modeling, and uncertainty in AI DC load profiles. Such frameworks are currently absent from the literature.

\subsection{Future Directions for AI Data Centers and Energy Storage}

\subsubsection{Power-Aware GPU Scheduling and Chip-Level ESS Co-Design}
A fundamental open problem is the co-design of GPU execution strategies with chip-level and rack-level ESSs. During training, it is essential to jointly optimize throughput and energy consumption through adaptive strategies for local and global batch size scheduling \cite{gu2304energy}. GPU resources should be dynamically scheduled across training tasks by accounting for time-varying electricity prices and renewable energy availability, enabling energy-aware model training \cite{chen2025electricity,wang2025providing,al2025instability,crozier2025potential}. Elastic resource allocation—dynamically adjusting the number of GPUs assigned to a training job—provides the controllability of GPU power consumption that is necessary for chip-level ESS or BBU scheduling. Beyond this, dynamic batch size selection can be implemented to regulate utility grid voltage profiles \cite{liang2026gpu}, directly linking compute scheduling to grid-level objectives. The integration of such power-aware scheduling with ESS dispatch optimization is an unexplored but promising research direction.

\subsubsection{AI DC Load Observability and Forecasting Infrastructure}
High-resolution monitoring of DC load dynamics requires waveform measurement units capable of capturing load profiles at very short time intervals (below 10 ms) \cite{chalamala2025data}. Such fine-grained measurements enable observation of fast transient behaviors that are invisible to conventional metering infrastructure. Building on high-fidelity data, machine learning techniques can predict short-term and long-term load patterns to support proactive ESS control and optimization \cite{lu2026dynamic,mughees2025short,jiang2025hyperload,mo2025learning}. Advanced load modeling studies further increase the load observability of AI DCs for both short-term operation and long-term planning. Close collaboration between utilities and DC operators is essential to enable secure, real-time access to load profiles, supporting coordinated grid–DC operation.

\subsubsection{Advanced Degradation Modeling, State Estimation, and RUL Prediction}
Lithium-ion batteries undergo both calendar and cyclic degradation \cite{Kostenko2024SecondLifeAging}. Degradation pathways including SEI growth, lithium plating, and particle fracture \cite{Hassan2023SecondLifeBatteries} lead to capacity fade and increased internal resistance. For SLBs in particular, aging curves are shaped by first-life usage patterns, making each battery unique. Estimating the remaining useful life (RUL) is essential for safe deployment, with approaches ranging from data-driven techniques \cite{Cui2025BMS2} and neural network-based models to entropy-based algorithms \cite{Strugnell2024EntropyBattery}. A degradation-aware dispatch strategy is needed to avoid pushing cells into the non-linear knee-point region. The EU Battery Passport initiative \cite{EU2023BatteryPassport} provides a lifecycle data framework that can inform second-life deployment decisions. Extending such frameworks to AI DC operating conditions, with their high cycling frequency and variable load profiles, is an important open research direction.

\subsubsection{FTM Grid-Scale BESS as a Reliable Reserve Asset}
By monitoring and communicating the grid-scale BESS status at AI DC sites with grid operators, more optimized reserve planning can be achieved \cite{7879307,8805438}. The total unused energy of the grid-scale BESS can be considered a potential reserve that can be injected into the grid during contingency events such as the tripping of synchronous generators. The FTM grid-scale BESS can also serve as an active reserve source in emergency cases by locally supplying the AI DC load in an islanded mode. This requires precise coordination and control strategies between utility and AI DC operators, as well as appropriate communication and dispatch protocols. Research on market mechanisms and regulatory frameworks that allow AI DCs to participate in reserve markets remains limited.

\subsubsection{Revisiting Solid-State Transformers (SSTs) for AI DCs}
SSTs \cite{8571261,sidorov2021solid,huang2016medium} are primarily considered for enabling medium-voltage AC to high-voltage DC conversion (e.g., 800~V DC or higher) to improve rack-level power density in AI DCs \cite{800vdc_nvidia}. In SST-based architectures, traditional UPS functionalities can be embedded within multi-stage power electronic converters, and the distribution transformer in the AI DC power delivery system can be eliminated, minimizing side effects of bidirectional power flow \cite{xu2026sequential}. However, SSTs face challenges including lower reliability due to multi-level power conversion stages, limited power ratings (typically a few MVA), and increased control complexity and cost. While attractive for high-density power delivery, SSTs are not yet suitable replacements for UPS systems at gigawatt-scale DCs. Multi-level modular SST architectures may improve scalability, but reliability and standardization remain key challenges requiring further investigation.

\subsubsection{Integrated AI DC Microgrid Design and Operation}
% An overarching future direction is the design and operation of AI DCs as active microgrids capable of seamlessly transitioning between grid-connected and islanded modes. Such AI DC microgrids would integrate GiUPS systems, grid-scale BESSs, FC backup generation, TES for cooling, and renewable on-site generation under a unified hierarchical EMS. The microgrid concept offers the potential for AI DCs to contribute positively to grid stability rather than being purely large and disruptive loads. Key open problems include optimal microgrid topology design for AI DC load characteristics, hierarchical control under uncertainty, and techno-economic co-optimization of all ESS layers jointly with compute scheduling.

AI DC microgrids are emerging as a promising architecture for improving both internal power resilience and external grid support \cite{irion2025optimizing,zhang2025unlocking}. These microgrids adopt hybrid power architectures that combine BESS, conventional generators, distributed generation units, and utility grid connections (see Fig. \ref{Fig_Datacenter_microgrid}) to achieve high reliability, redundancy, and operational flexibility. Within this framework, the placement and coordination of BESS are critical design choices, since properly located BESS can damp fast AI workload-induced load fluctuations, enhance local voltage and frequency stability, and enable coordinated operation among diverse power generation units. 

In addition to supporting internal reliability, BESS can also provide grid services such as demand response, peak shaving, and fast reserves. This broader functionality increasingly blurs the traditional boundary between backup-oriented systems and active grid-support assets, particularly in relation to GiUPS systems, and raises new questions regarding their joint design and control. More broadly, the integration of GiUPS systems, grid-scale BESSs, FC backup generation, TES for cooling, and renewable on-site generation under a unified hierarchical EMS points toward a future in which AI DCs operate as active microgrids capable of seamless transitions between grid-connected and islanded modes. In this vision, AI DCs would no longer behave merely as large and potentially disruptive loads, but rather as flexible grid participants that can reshape power interface thresholds and redefine power quality metrics. However, several open challenges remain, including optimal microgrid topology design for AI DC load characteristics, hierarchical control under uncertainty, and the techno-economic co-optimization of all ESS layers jointly with compute scheduling.

{\begin{figure}[t]
\centering
\includegraphics[width=0.95\linewidth]{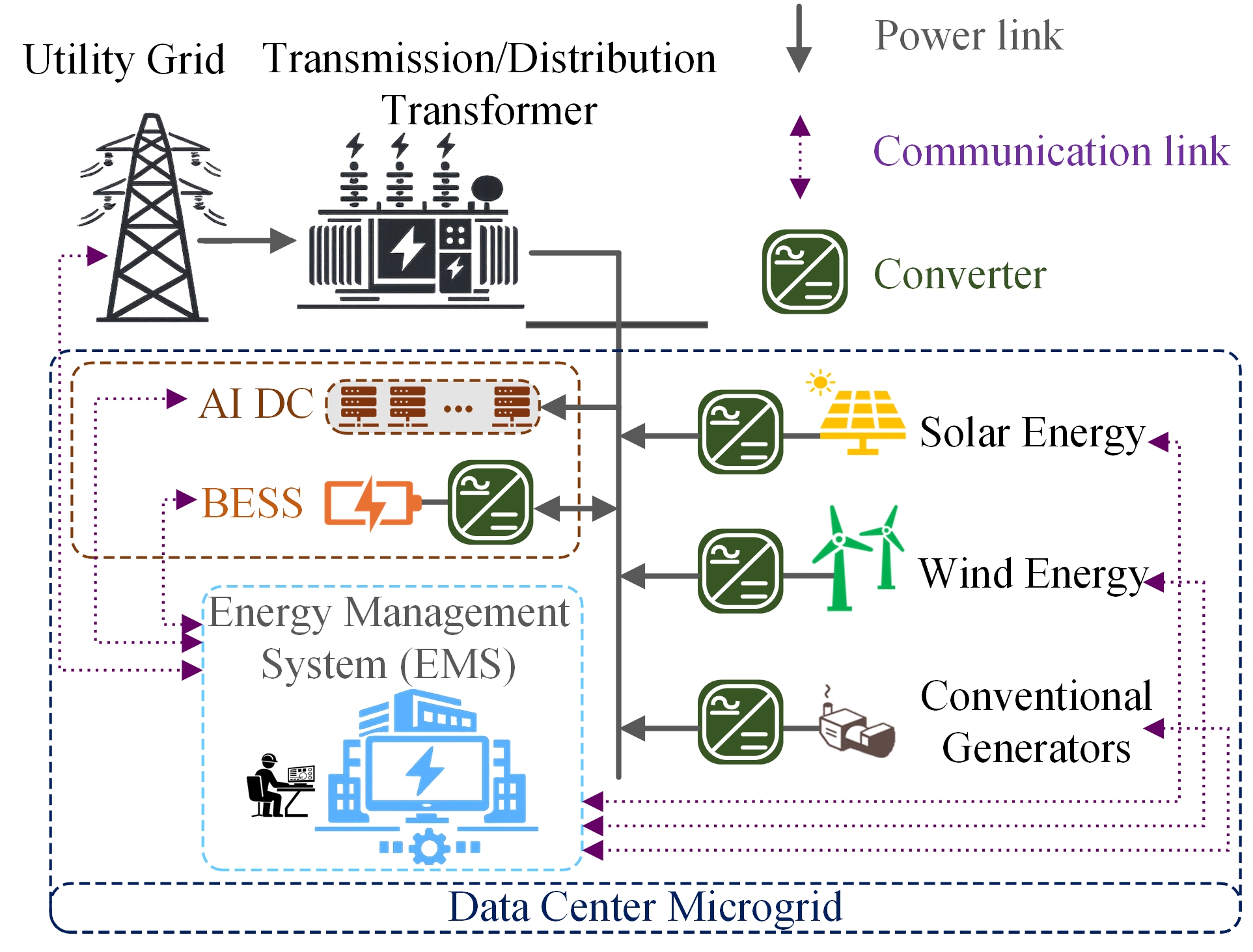}
\caption{AI DC microgrid.}
\label{Fig_Datacenter_microgrid}
\vspace{-5mm}
\end{figure}}

% \vspace{-5mm}
\section{Conclusions} \label{conclusion}

This paper has provided a comprehensive critical review of multi-layer ESSs for AI DCs and their roles in supporting both facility-level reliability and utility grid integration. Organized around a four-layer hierarchical taxonomy, namely chip-level buffers, rack/server-level ESSs, facility-level GiUPS systems, and grid-scale BESSs, the review systematically analyzes each technology's response timescale, application role, advantages, limitations, and coordination requirements.

Several key findings emerge from this review. First, AI DC load profiles are fundamentally different from traditional grid loads. Sub-second power variability, high peak-to-average ratios, and unpredictable workload transitions make conventional ESS dispatch strategies insufficient. Second, GiUPS systems represent the most significant architectural evolution in DC power infrastructure, transforming passive UPS batteries into active grid assets capable of FFR, VRT, peak shaving, and synthetic inertia provision. Third, grid-scale BESSs, whether BTM or FTM, are essential for demand charge reduction, renewable integration, and ancillary service provision at the facility level, but their sizing must explicitly account for degradation, load uncertainty, and multi-objective operational requirements. Fourth, non-battery technologies, namely FCs and TES, are underutilized in the current literature relative to their potential. FCs offer a viable clean replacement for diesel backup generation, while TES enables significant cooling efficiency gains and RES utilization. Fifth, SLBESSs offer a cost-effective deployment pathway for large-scale storage in AI DCs, provided that SoH uncertainty and cell heterogeneity are addressed through rigorous degradation-aware control.

The review identifies eight critical open challenges: the absence of multi-physics simulation tools for AI DC EMS validation, the immaturity of energy-aware GPU scheduling, bidirectional power flow impacts on distribution transformers, insufficient load forecasting methods for AI workloads, inadequate degradation models under AI DC cycling conditions, the lack of hierarchical ESS coordination frameworks, LDES technology immaturity, and the absence of multi-layer techno-economic sizing methodologies. Future research should address these challenges through the co-design of compute scheduling and ESS dispatch, the development of high-resolution load monitoring infrastructure, advanced degradation and RUL estimation methods for AI DC operating conditions, and integrated AI DC microgrid architectures. Addressing these open problems will be essential for transforming AI DCs from large disruptive grid loads into flexible, grid-supportive assets that contribute to power system reliability and the sustainable integration of renewable energy.

% \vspace{-5mm}

\bibliographystyle{IEEEtran}
\bibliography{Ref}

@techreport{chalamala2025data,
  author      = {Chalamala, Babu and others},
  title       = {Data Center Growth and Grid Readiness (TR131)},
  institution = {IEEE Power and Energy Society},
  year        = {2025},
  number      = {TR131},
  type        = {Technical Report},
  doi         = {10.17023/w4wy-s557}
}

@techreport{CharacteristicsandRisksofEmergingLargeLoads,
  title        = {Characteristics and Risks of Emerging Large Loads},
  author       = {{North American Electric Reliability Corporation (NERC)}},
  institution  = {NERC},
  year         = {2025},
  month        = {July},
  type         = {White Paper},
  url          = {https://www.nerc.com/globalassets/who-we-are/standing-committees/rstc/whitepaper-characteristics-and-risks-of-emerging-large-loads.pdf},
  note         = {Large Loads Task Force (LLTF)}
}

@misc{VirginiaDataCenter,
  title        = {Big Tech’s data center boom poses new risk to US grid operators},
  author       = {Tim McLaughlin},
  year         = {2025},
  url          = {https://www.reuters.com/technology/big-techs-data-center-boom-poses-new-risk-us-grid-operators-2025-03-19/},
}

@misc{WhatisanAIdatacenter,
  title        = {What is an AI data center?},
  author={Alexandra Jonker and Alice Gomstyn},
  year         = {2025},
  url          = {https://www.ibm.com/think/topics/ai-data-center},
}

@techreport{DataCenterDevelopmentinanAI-DrivenMarket,
  author      = {{Stream Data Centers}},
  title       = {Data Center Development in an AI-Driven Market},
  institution = {Stream Data Centers},
  year        = {2024},
  month       = feb,
  url         = {https://www.streamdatacenters.com/wp-content/uploads/2024/02/SDC-BTPS-Whitepaper-240222.pdf}
}

@article{SAFARI2026120386,
title = {A research-industry perspective of battery systems technology for next-generation data centers},
journal = {Journal of Energy Storage},
volume = {152},
pages = {120386},
year = {2026},
issn = {2352-152X},
doi = {https://doi.org/10.1016/j.est.2026.120386},
url = {https://www.sciencedirect.com/science/article/pii/S2352152X26000502},
author = {Ashkan Safari and Frede Blaabjerg and Arman Oshnoei}
}

@online{AnOverviewofPopularNVIDIAGPUs,
  author       = {{Atlantic.Net}},
  title        = {An Overview of Popular NVIDIA GPUs},
  year         = {2025},
  url          = {https://www.atlantic.net/gpu-server-hosting/an-overview-of-popular-nvidia-gpus/},
  organization = {Atlantic.Net},
  note         = {Accessed: 2026-02-27}
}

@online{HowNewGB300NVL72FeaturesProvideSteadyPowerforAI,
  author       = {Rouslan Dimitrov, Harry Petty, Neeraj Srivastava and Mathias Blake},
  title        = {How New GB300 NVL72 Features Provide Steady Power for AI},
  year         = {2025},
  organization = {NVIDIA Developer Blog},
  url          = {https://developer.nvidia.com/blog/how-new-gb300-nvl72-features-provide-steady-power-for-ai/}
}

@online{AccelerateAIInferencewithGoogleCloudTPUsandGPUs,
  author       = {{Google Cloud}},
  title        = {Accelerating AI Inference with Google Cloud TPUs and GPUs},
  year         = {2024},
  day          = {10},
  url          = {https://cloud.google.com/blog/products/compute/accelerating-ai-inference-with-google-cloud-tpus-and-gpus},
  organization = {Google Cloud Blog},
  note         = {Accessed: 2026-02-27}
}

@online{WhatchangesinstoragewillAIdrive?,
  author       = {{Micron Technology}},
  title        = {What Changes in Storage Will AI Drive?},
  year         = {2024},
  url          = {https://www.micron.com/about/blog/storage/ai/what-changes-in-storage-will-ai-drive},
  organization = {Micron Blog}
}

@online{GPUDirectStorage:ADirectPathBetweenStorageandGPUMemory,
  author       = {{NVIDIA Corporation}},
  title        = {GPUDirect Storage: A Direct Path Between Storage and GPU Memory},
  year         = {2019},
  url          = {https://developer.nvidia.com/blog/gpudirect-storage/},
  organization = {NVIDIA Developer Blog}
}

@online{NetworkingforDataCentersandtheEraofAI,
  author       = {{NVIDIA Corporation}},
  title        = {Networking for Data Centers and the Era of AI},
  year         = {2023},
  url          = {https://developer.nvidia.com/blog/networking-for-data-centers-and-the-era-of-ai/},
  organization = {NVIDIA Developer Blog}
}

@inproceedings{alibaba-hpn,
  author    = {Kun Qian and Yongqing Xi and Jiamin Cao and Jiaqi Gao and others},
  title     = {Alibaba HPN: A Data Center Network for Large Language Model Training},
  booktitle = {Proceedings of the ACM SIGCOMM 2024 Conference},
  year      = {2024},
  pages     = {691--706},
  publisher = {Association for Computing Machinery},
  address   = {New York, NY, USA},
  doi       = {10.1145/3651890.3672265},
  url       = {https://dl.acm.org/doi/10.1145/3651890.3672265}
}

@inproceedings{meta_roce,
author = {Gangidi, Adithya and Miao, Rui and Zheng, Shengbao and Bondu, Sai Jayesh and Goes, Guilherme and Morsy, Hany and Puri, Rohit and Riftadi, Mohammad and Shetty, Ashmitha Jeevaraj and Yang, Jingyi and Zhang, Shuqiang and Fernandez, Mikel Jimenez and Gandham, Shashidhar and Zeng, Hongyi},
title = {RDMA over Ethernet for Distributed Training at Meta Scale},
year = {2024},
isbn = {9798400706141},
publisher = {Association for Computing Machinery},
address = {New York, NY, USA},
url = {https://doi.org/10.1145/3651890.3672233},
doi = {10.1145/3651890.3672233},
booktitle = {Proceedings of the ACM SIGCOMM 2024 Conference},
pages = {57–70},
numpages = {14},
location = {Sydney, NSW, Australia},
series = {ACM SIGCOMM '24}
}

@online{GPU-to-GPUCommunication:UnlockingParallelismBeyondtheCore,
  author       = {Nikheel Vishwas Savant},
  title        = {GPU-to-GPU Communication: Unlocking Parallelism Beyond the Core},
  year         = {2025},
  organization = {Medium},
  url          = {https://medium.com/@nikheelvs/gpu-to-gpu-communication-unlocking-parallelism-beyond-the-core-a80de2974078},
}

@ARTICLE{10255526,
  author={Sheehan, Stuart and Rakow, Alexander},
  journal={IEEE Electrification Magazine}, 
  title={Evolving a Data Center Into a Microgrid: Industry perspectives and lessons learned}, 
  year={2023},
  volume={11},
  number={3},
  pages={16-25},
  doi={10.1109/MELE.2023.3291193}}

@online{AdvantagesandChallengesofNuclear-PoweredDataCenters,
  author       = {{U.S. Department of Energy}},
  title        = {Advantages and Challenges of Nuclear-Powered Data Centers},
  year         = {2025},
  organization = {Office of Nuclear Energy},
  url          = {https://www.energy.gov/ne/articles/advantages-and-challenges-nuclear-powered-data-centers},
  note         = {Accessed: 2026-02-27}
}

@ARTICLE{11230048,
  author={Nehrir, Hashem and Wang, Caisheng},
  journal={IEEE Energy Sustainability Magazine}, 
  title={Hydrogen Fuel and Fuel Cells: Potential Candidates for Sustainable, Dispatchable, and Environmentally Friendly Power Generation and Transport Technologies}, 
  year={2025},
  volume={1},
  number={3},
  pages={52-65},
  doi={10.1109/ESM.2025.3606163}}

@online{Understandingdirect-to-chipcoolinginHPCinfrastructure:Adeepdiveintoliquidcooling,
  author       = {{Vertiv}},
  title        = {Understanding Direct-to-Chip Cooling in HPC Infrastructure: A Deep Dive into Liquid Cooling},
  year         = {2024},
  organization = {Vertiv},
  url          = {https://www.vertiv.com/en-us/about/news-and-insights/articles/educational-articles/understanding-direct-to-chip-cooling-in-hpc-infrastructure-a-deep-dive-into-liquid-cooling/},
}

@article{rahman2026energy,
  title={Energy Storage Systems for AI Data Centers: A Review of Technologies, Characteristics, and Applicability},
  author={Rahman, Saifur and Khan, Tafsir Ahmed},
  journal={Energies},
  volume={19},
  number={3},
  pages={634},
  year={2026},
  publisher={MDPI}
}

@article{ginzburg2025technical,
  title={Technical Challenges of AI Data Center Integration into Power Grids—A Survey},
  author={Ginzburg-Ganz, Elinor and Lifshits, Pavel and Machlev, Ram and Belikov, Juri and Krieger, Ziv and Levron, Yoash},
  journal={Energies},
  volume={19},
  number={1},
  pages={137},
  year={2025},
  publisher={MDPI}
}

@article{chen2025electricity,
  title={Electricity demand and grid impacts of AI data centers: Challenges and prospects},
  author={Chen, Xin and Wang, Xiaoyang and Colacelli, Ana and Lee, Matt and Xie, Le},
  journal={arXiv preprint arXiv:2509.07218},
  year={2025}
}

@article{lu2026dynamic,
  title={Dynamic Load Model for Data Centers with Pattern-Consistent Calibration},
  author={Lu, Siyu and Xiao, Chenhan and Weng, Yang},
  journal={arXiv preprint arXiv:2602.07859},
  year={2026}
}

@article{vercellino2026measurement,
  title={Measurement of Generative AI Workload Power Profiles for Whole-Facility Data Center Infrastructure Planning},
  author={Vercellino, Roberto and Willard, Jared and Campos, Gustavo and Pereira, Weslley da Silva and Hull, Olivia and Selensky, Matthew and Mueller, Juliane},
  journal={arXiv preprint arXiv:2604.07345},
  year={2026}
}

@article{xu2026sequential,
  title={Sequential Operating Simulation of Solid State Transformer-Driven Next-Generation 800 VDC Data Center},
  author={Xu, Jian and Jiang, Xinxiong and Bao, Yi and Zheng, Yuchen and Chen, Xuhui and Xu, Qiang and Liao, Siyang and Ke, Deping and Gao, Xiaoqi},
  journal={arXiv preprint arXiv:2601.16502},
  year={2026}
}

@ARTICLE{11369875,
  author={Li, Xiang and Guan, Yueshi and Yao, Tingting and Wang, Yijie and Wang, Wei and Xu, Dianguo},
  journal={IEEE Transactions on Power Electronics}, 
  title={A Soft-Switching Multiresonant Switched-Capacitor Converter for Data Center Applications}, 
  year={2026},
  volume={41},
  number={7},
  pages={11079-11097},
  doi={10.1109/TPEL.2026.3657465}}

@article{choukse2025power,
  title={Power stabilization for AI training datacenters},
  author={Choukse, Esha and Warrier, Brijesh and Heath, Scot and Belmont, Luz and Zhao, April and Khan, Hassan Ali and Harry, Brian and Kappel, Matthew and Hewett, Russell J and Datta, Kushal and others},
  journal={arXiv preprint arXiv:2508.14318},
  year={2025}
}

@misc{Eaton2025DataCentersGoodGridCitizen,
  author       = {Watson, Keith},
  title        = {Data Centers -- A Good Grid Citizen},
  howpublished = {Presentation},
  organization = {Eaton},
  year         = {2025},
  month        = jul,
  day          = {11},
  note         = {Slide on grid support and batteries; Mission Critical Solutions presentation}
}

@techreport{SchneiderWP185BESS,
  author       = {Donovan, Patrick},
  title        = {Understanding BESS: Battery Energy Storage Systems for Data Centers},
  institution  = {Schneider Electric Energy Management Research Center},
  number       = {White Paper 185},
  version      = {1.1},
  year         = {2025}
}

@techreport{Vertiv2025GridInteractiveUPS,
  author       = {Di Filippi, Arturo and Valentini, Luca},
  title        = {How to Maximize Revenues from Your Data Center Energy Storage System with Grid Interactive UPS},
  institution  = {Vertiv},
  year         = {2025},
  note         = {Vertiv White Paper}
}

@article{mamun2016multi,
  title={Multi-objective optimization of demand response in a datacenter with lithium-ion battery storage},
  author={Mamun, A and Narayanan, I and Wang, D and Sivasubramaniam, A and Fathy, HK},
  journal={Journal of Energy Storage},
  volume={7},
  pages={258--269},
  year={2016},
  publisher={Elsevier}
}

@article{ko2025mitigation,
  title={Mitigation of Datacenter Demand Ramping and Fluctuation using Hybrid ESS and Supercapacitor},
  author={Ko, Min-Seung and Shim, Jae Woong and Zhu, Hao},
  journal={arXiv preprint arXiv:2512.08076},
  year={2025}
}

@techreport{NERC2025LargeLoads,
  author       = {{North American Electric Reliability Corporation (NERC)}},
  title        = {Characteristics and Risks of Emerging Large Loads},
  institution  = {North American Electric Reliability Corporation},
  note         = {Large Loads Task Force White Paper},
  month        = jul,
  year         = {2025}
}

@techreport{NERC2023GridFormingBESS,
  author       = {{North American Electric Reliability Corporation (NERC)}},
  title        = {Grid Forming Functional Specifications for BPS-Connected Battery Energy Storage Systems},
  institution  = {North American Electric Reliability Corporation},
  year         = {2023},
  month        = sep,
  type         = {White Paper},
  url          = {https://www.nerc.com/globalassets/our-work/reports/white-papers/white_paper_gfm_functional_specification.pdf},
}

@inproceedings{ahrabi2025ai,
  title={AI-Driven Data Center Energy Profile, Power Quality, Sustainable Sitting, and Energy Management: A Comprehensive Survey},
  author={Ahrabi, Rouzbeh Reza and Mousavi, Alireza and Mohammadi, Ebrahim and Wu, Ryan and Chen, Aoxia Kevin},
  booktitle={2025 IEEE Conference on Technologies for Sustainability (SusTech)},
  pages={1--8},
  year={2025},
  organization={IEEE}
}

@INPROCEEDINGS{8217257,
  author={Cupelli, Lisette and Barve, Nikhil and Monti, Antonello},
  booktitle={IECON 2017 - 43rd Annual Conference of the IEEE Industrial Electronics Society}, 
  title={Optimal sizing of data center battery energy storage system for provision of frequency containment reserve}, 
  year={2017},
  volume={},
  number={},
  pages={7185-7190},
  doi={10.1109/IECON.2017.8217257}}

@article{Hassan2023SecondLifeBatteries,
  author  = {Hassan, A. and Khan, S. A. and Li, R. and Su, W. and Zhou, X. and Wang, M. and Wang, B.},
  title   = {Second-Life Batteries: A Review on Power Grid Applications, Degradation Mechanisms, and Power Electronics Interface Architectures},
  journal = {Batteries},
  volume  = {9},
  number  = {12},
  pages   = {571},
  year    = {2023}
}

@misc{B2UStorage,
  author = {{B2U Storage Solutions}},
  title  = {B2U Storage Solutions},
  year   = {2025},
  url    = {https://www.b2uco.com/}
}

@misc{ElementEnergy,
  author = {{Element Energy}},
  title  = {Element Energy},
  year   = {2025},
  url    = {https://elementenergy.com/}
}

@misc{MomentEnergy,
  author = {{Moment Energy}},
  title  = {Moment Energy},
  year   = {2025},
  url    = {https://www.momentenergy.com/}
}

@misc{Smartville,
  author = {{Smartville}},
  title  = {Smartville},
  year   = {2025},
  url    = {https://smartville.io/}
}

@misc{Redwood2025AIStorage,
  author = {{Redwood Energy}},
  title  = {Fast, Low-Cost Storage to Power the Age of AI and a Changing Grid},
  year   = {2025},
  url    = {https://www.redwoodmaterials.com/news/redwood-energy-fast-low-cost-storage-to-power-the-age-of-ai-and-a-changing-grid/}
}

@article{Bobba2018SecondLifeLCA,
  author  = {Bobba, S. et al.},
  title   = {Life Cycle Assessment of Repurposed Electric Vehicle Batteries: Environmental and Climate Benefits of Second-Life Use},
  journal = {Journal of Energy Storage},
  volume  = {19},
  pages   = {213--225},
  year    = {2018}
}

@misc{CrusoeRedwood2025Microgrid,
  author = {{Crusoe Energy} and {Redwood Materials}},
  title  = {Crusoe and Redwood Materials Power AI/Data Center Infrastructure Using Second-Life EV Battery Microgrid},
  year   = {2025},
  month  = jun,
  url    = {https://www.crusoe.ai/resources/newsroom/crusoe-and-redwood-materials-power-the-future-of-ai}
}

@article{Wang2023LifeExtendedControl,
  author  = {Wang, H. and Rasheed, M. and Hassan, R. and Kamel, M. and Tong, S. and Zane, R.},
  title   = {Life-Extended Active Battery Control for Energy Storage Using Electric Vehicle Retired Batteries},
  journal = {IEEE Transactions on Power Electronics},
  volume  = {38},
  number  = {6},
  pages   = {6801--6805},
  year    = {2023}
}

@misc{Kostenko2024SecondLifeAging,
  author = {Kostenko, G. P.},
  title  = {Accounting Calendar and Cyclic Ageing Factors in Diagnostic and Prognostic Models of Second-Life EV Batteries},
  year   = {2024}
}

@article{Cui2025BMS2,
  author  = {Cui, X. and Khan, M. A. and Singh, S. and Sharma, R. and Onori, S.},
  title   = {Toward a BMS 2.0 Design Framework: Adaptive Data-Driven State-of-Health Estimation for Second-Life Batteries with BIBO Stability Guarantees},
  journal = {IEEE Transactions on Transportation Electrification},
  year    = {2025}
}

@misc{Strugnell2024EntropyBattery,
  author = {Strugnell-Lees, B. and Evdokimova, E. and Wik, T.},
  title  = {An Entropy-Based, Self-Adaptive Predictive Algorithm for Battery Degradation},
  year   = {2024}
}

@misc{EU2023BatteryPassport,
  author = {{European Union}},
  title  = {Regulation (EU) 2023/1542},
  year   = {2023},
  url    = {https://eur-lex.europa.eu/eli/reg/2023/1542/oj}
}

@misc{BNEF2025BatteryCost,
  author = {{BloombergNEF}},
  title  = {Lithium-Ion Battery Pack Prices Fall to \$108 per Kilowatt-Hour Despite Rising Metal Prices},
  year   = {2025},
  url    = {https://about.bnef.com/insights/clean-transport/lithium-ion-battery-pack-prices-fall-to-108-per-kilowatt-hour-despite-rising-metal-prices-bloombergnef/}
}

@techreport{Bach2025SecondLifeValue,
  author      = {Bach, A. and Onori, S. and Reichelstein, S. and Zhuang, J.},
  title       = {Fair Market Value of Used Capacity Assets: Forecasts for Repurposed Electric Vehicle Batteries},
  institution = {ZEW -- Centre for European Economic Research},
  number      = {Discussion Paper No. 25-065},
  year        = {2025}
}

@misc{UL1974,
  author = {{UL}},
  title  = {UL 1974: Evaluation for Repurposing Batteries},
  year   = {2024},
  url    = {https://www.ul.com/services/second-life-electric-vehicle-battery-repurposing-facility-certification}
}

@misc{IRSOBBB,
  author = {{Internal Revenue Service (IRS)}},
  title  = {Domestic Content Bonus Credit},
  year   = {2025},
  url    = {https://www.irs.gov/credits-deductions/clean-electricity-production-credit}
}

@article{pan2025salt,
  title={Salt cavern redox flow battery: The next-generation long-duration, large-scale energy storage system},
  author={Pan, Lyuming and Song, Manrong and Muzaffar, Nimra and Chen, Liuping and Ji, Chao and Yao, Shengxin and Xu, Junhui and Wu, Weixiong and Li, Yubai and Chen, Jie and others},
  journal={Current Opinion in Electrochemistry},
  volume={49},
  pages={101604},
  year={2025},
  publisher={Elsevier}
}

@article{tao2025coordinated,
  title={Coordinated Fast Frequency Response from Electric Vehicles, Data Centers, and Battery Energy Storage Systems},
  author={Tao, Xiaojie and Gadh, Rajit},
  journal={arXiv preprint arXiv:2512.14136},
  year={2025}
}

@article{ye2026grid,
  title={A Grid-Forming Energy Storage System Capacity Planning Method Considering Device Lifetime},
  author={Ye, Guisen and Fang, Jingyang and Wang, Nan and Gaogao, Yinan and Sun, Kangyuan},
  journal={Energies},
  year={2026},
  publisher={Multidisciplinary Digital Publishing Institute}
}

@article{yu2025reliability,
  title={Reliability and economic impacts of utilizing battery energy storage in data centers for energy flexibility services in smart grids},
  author={Yu, Yang and Shan, Kui and Tang, Hong and Wang, Shengwei},
  journal={Energy Conversion and Management},
  volume={339},
  pages={119951},
  year={2025},
  publisher={Elsevier}
}

@misc{QuantaTech2025AILoadProfiles,
  author       = {{Quanta Tech LLC}},
  title        = {Understanding AI Load Profiles and Their Impact on Power Systems},
  howpublished = {Online Seminar (Webinar)},
  year         = {2025},
  month        = sep,
  day          = {3},
}

@article{casals2019second,
  title={Second life batteries lifespan: Rest of useful life and environmental analysis},
  author={Casals, Lluc Canals and Garc{\'\i}a, B Amante and Canal, Camille},
  journal={Journal of environmental management},
  volume={232},
  pages={354--363},
  year={2019},
  publisher={Elsevier}
}

@article{martinez2018battery,
  title={Battery second life: Hype, hope or reality? A critical review of the state of the art},
  author={Martinez-Laserna, Egoitz and Gandiaga, I{\~n}igo and Sarasketa-Zabala, Elixabet and Badeda, Julia and Stroe, D-I and Swierczynski, Maciej and Goikoetxea, Ander},
  journal={Renewable and Sustainable Energy Reviews},
  volume={93},
  pages={701--718},
  year={2018},
  publisher={Elsevier}
}

@article{neubauer2011ability,
  title={The ability of battery second use strategies to impact plug-in electric vehicle prices and serve utility energy storage applications},
  author={Neubauer, Jeremy and Pesaran, Ahmad},
  journal={Journal of Power Sources},
  volume={196},
  number={23},
  pages={10351--10358},
  year={2011},
  publisher={Elsevier}
}

@article{vetter2005ageing,
  title={Ageing mechanisms in lithium-ion batteries},
  author={Vetter, Jens and Nov{\'a}k, Petr and Wagner, Markus Robert and Veit, Claudia and M{\"o}ller, K-C and Besenhard, JO and Winter, Martin and Wohlfahrt-Mehrens, Margret and Vogler, Christoph and Hammouche, Abderrezak},
  journal={Journal of power sources},
  volume={147},
  number={1-2},
  pages={269--281},
  year={2005},
  publisher={Elsevier}
}

@phdthesis{perez2025enabling,
  title={Enabling Net Zero Data Centers: A Techno-economic Analysis of Bloom Energy’s SOFC Systems},
  author={Perez, Felipe},
  year={2025},
  school={Politecnico di Torino}
}

@article{fan2021recent,
  title={Recent development of hydrogen and fuel cell technologies: A review},
  author={Fan, Lixin and Tu, Zhengkai and Chan, Siew Hwa},
  journal={Energy Reports},
  volume={7},
  pages={8421--8446},
  year={2021},
  publisher={Elsevier}
}

@article{kong2025grid,
  title={A Grid-Forming Control Method for PEMFC Power Conversion Systems with Power Ramp Rate Limitation to Prevent Fuel Starvation},
  author={Kong, Im-Bo and Kim, Wook-Sung and Chae, Suyong},
  journal={IEEE Open Journal of Power Electronics},
  year={2025},
  publisher={IEEE}
}

@article{nikiforakis2025understanding,
  title={Understanding solid oxide fuel cell hybridization: a critical review},
  author={Nikiforakis, Ioannis and Mamalis, Sotirios and Assanis, Dimitris},
  journal={Applied Energy},
  volume={377},
  pages={124277},
  year={2025},
  publisher={Elsevier}
}

@article{manzo2025fuel,
  title={Fuel cell technology review: Types, economy, applications, and vehicle-to-grid scheme},
  author={Manzo, Danny and Thai, Ryan and Le, Ha Thu and Venayagamoorthy, Ganesh Kumar},
  journal={Sustainable Energy Technologies and Assessments},
  volume={75},
  pages={104229},
  year={2025},
  publisher={Elsevier}
}

@article{nikiforow2018power,
  title={Power ramp rate capabilities of a 5 kW proton exchange membrane fuel cell system with discrete ejector control},
  author={Nikiforow, Kaj and Pennanen, Jari and Ihonen, Jari and Uski, Sanna and Koski, Pauli},
  journal={Journal of Power Sources},
  volume={381},
  pages={30--37},
  year={2018},
  publisher={Elsevier}
}

@article{li2025ai,
  title={AI Load Dynamics--A Power Electronics Perspective},
  author={Li, Yuzhuo and Li, Yunwei},
  journal={arXiv preprint arXiv:2502.01647},
  year={2025}
}

@article{zheng2025techno,
  title={Techno-economic assessment framework for 2.5 MW-scale grid-connected proton exchange membrane fuel cell power systems: A case study in China},
  author={Zheng, Peijun and Xie, Xiaorong and Zhang, Chunpeng and Cai, Shanshan and Pan, Jun and Zhang, Hang and Yan, Mingyu and Mu, Qing},
  journal={International Journal of Hydrogen Energy},
  volume={167},
  pages={150942},
  year={2025},
  publisher={Elsevier}
}

@article{enescu2020thermal,
  title={Thermal energy storage for grid applications: Current status and emerging trends},
  author={Enescu, Diana and Chicco, Gianfranco and Porumb, Radu and Seritan, George},
  journal={Energies},
  volume={13},
  number={2},
  pages={340},
  year={2020},
  publisher={MDPI}
}

@article{sarbu2018comprehensive,
  title={A comprehensive review of thermal energy storage},
  author={Sarbu, Ioan and Sebarchievici, Calin},
  journal={Sustainability},
  volume={10},
  number={1},
  pages={191},
  year={2018},
  publisher={MDPI}
}

@article{alva2018overview,
  title={An overview of thermal energy storage systems},
  author={Alva, Guruprasad and Lin, Yaxue and Fang, Guiyin},
  journal={Energy},
  volume={144},
  pages={341--378},
  year={2018},
  publisher={Elsevier}
}

@article{guelpa2019thermal,
  title={Thermal energy storage in district heating and cooling systems: A review},
  author={Guelpa, Elisa and Verda, Vittorio},
  journal={Applied Energy},
  volume={252},
  pages={113474},
  year={2019},
  publisher={Elsevier}
}

@article{luerssen2020life,
  title={Life cycle cost analysis (LCCA) of PV-powered cooling systems with thermal energy and battery storage for off-grid applications},
  author={Luerssen, Christoph and Gandhi, Oktoviano and Reindl, Thomas and Sekhar, Chandra and Cheong, David},
  journal={Applied Energy},
  volume={273},
  pages={115145},
  year={2020},
  publisher={Elsevier}
}

@article{omrani2025ai,
  title={AI-driven optimization of fan-wall cooling system in a medium-density data center},
  author={Omrani, Mostafa and Ghassemi, Majid},
  journal={International Journal of Heat and Mass Transfer},
  volume={247},
  pages={127159},
  year={2025},
  publisher={Elsevier}
}

@inproceedings{karpati2012uninterruptible,
  title={Uninterruptible Power Supplies (UPS) for data center},
  author={Karpati, A and Zsigmond, Gy and V{\"o}r{\"o}s, M and Lendvay, Marianna},
  booktitle={2012 IEEE 10th Jubilee International Symposium on Intelligent Systems and Informatics},
  pages={351--355},
  year={2012},
  organization={IEEE}
}

@article{fawaz2019minimizing,
  title={Minimizing data center uninterruptable power supply overload by server power capping},
  author={Fawaz, AL-Hazemi and Lorincz, Josip and Mohammed, Alaelddin FY},
  journal={IEEE Communications Letters},
  volume={23},
  number={8},
  pages={1342--1346},
  year={2019},
  publisher={IEEE}
}

@article{wang2025coordinated,
  title={Coordinated optimization of distributed energy system and storage-enhanced uninterruptible power supply in data center: A three-level optimization framework with model predictive control},
  author={Wang, Zimu and Yin, Zhiqiang and Yang, Jinyu and Wang, Jiangjiang},
  journal={Energy Conversion and Management},
  volume={342},
  pages={120137},
  year={2025},
  publisher={Elsevier}
}

@inproceedings{sidorov2021solid,
  title={Solid state transformer as a part electrical circuit for data center application},
  author={Sidorov, Andrey and Zinoviev, Gennady and Petzoldt, J{\"u}rgen},
  booktitle={2021 IEEE 22nd International Conference of Young Professionals in Electron Devices and Materials (EDM)},
  pages={354--359},
  year={2021},
  organization={IEEE}
}

@article{huang2016medium,
  title={Medium-voltage solid-state transformer: Technology for a smarter and resilient grid},
  author={Huang, Alex Q},
  journal={IEEE Industrial Electronics Magazine},
  volume={10},
  number={3},
  pages={29--42},
  year={2016},
  publisher={IEEE}
}

@misc{MaximizeRevenues,
  author = {Di Filippi, Arturo and Valentini, Luca },
  title  = {How to Maximize Revenues from Your Data Center Energy Storage System with Grid Interactive UPS},
  year   = {2025},
  url    = {https://www.vertiv.com/4918e5/globalassets/documents/white-papers/white-paper-maximize-revenues-data-center-energy-storage-grid-ups_329440_2.pdf}
}

@misc{VoltageRide-Through,
  author = {Christopher Tozzi},
  title  = {Voltage Ride-Through: A Key Ingredient in Data Center Resilience},
  year   = {2026},
  url    = {https://www.datacenterknowledge.com/uptime/voltage-ride-through-a-key-ingredient-in-data-center-resilience}
}

@misc{MediumVoltageUPS,
  author = {ON.energy},
  title  = {Hyperscale Meets Grid Stability: ON.energy Launches Medium-Voltage UPS Built for AI Data Centers},
  year   = {2025},
  url    = {https://www.nacleanenergy.com/alternative-energies/hyperscale-meets-grid-stability-on-energy-launches-medium-voltage-ups-built-for-ai-data-centers}
}

@article{xie2025enhancing,
  title={Enhancing Data Center Low-Voltage Ride-Through},
  author={Xie, Yiheng and Cui, Wenqi and Wierman, Adam},
  journal={arXiv preprint arXiv:2510.03867},
  year={2025}
}

@ARTICLE{10255103,
  author={Paananen, Janne},
  journal={IEEE Electrification Magazine}, 
  title={Grid-Interactive Data Centers Enabling Energy Transition: Data center’s hidden potential to provide essential grid services of a future power system}, 
  year={2023},
  volume={11},
  number={3},
  pages={26-34},
  doi={10.1109/MELE.2023.3291195}}

@misc{800vdc_nvidia,
  title        = {800 VDC Architecture for Next-Generation AI Infrastructure},
  author       = {Jared Huntington and Mike Tu},
  year         = {2025},
  url = {https://nvdam.nvidia.com/assets/share/asset/zlg5snufeo}
}

@misc{ocp_orv3_bbu_module_spec_Rev1.4,
  title        = {Open Compute Project: Open Rack V3 48V BBU (Rev 1.4)},
  author       = {Sun, David and Shapiro, Dmitriy and Kim, Ben and Athavale, Jayati and Mercado, Rommel},
  url = {https://www.opencompute.org/documents/open-rack-v3-bbu-module-spec-1-4-pdf},
  year         = {2023},
}

@misc{ocp_orv3_bbu_shelf_spec_Rev1.1,
  title        = {Open Compute Project: Open Rack V3 BBU Shelf (Rev 1.1)},
  author       = {.Sun, David and Shapiro, Dmitriy and Kim, Ben and Athavale, Jayati and Mercado, Rommel},
  url = {https://www.opencompute.org/documents/open-rack-v3-bbu-shelf-spec-rev1-1-pdf-1},
  year         = {2022},
}

@misc{adi_ocp_orv3_bbu_reference,
  title        = {ADI OCP ORV3 BBU Reference Design},
  author       = {{Analog Devices}},
  url = {https://wiki.analog.com/resources/eval/adi-ocp-orv3-bbu-reference-design },
  year         = {2024}
}

@inproceedings{micro20_priority_charging,
  author={Malla, Sulav and Deng, Qingyuan and Ebrahimzadeh, Zoh and Gasperetti, Joe and Jain, Sajal and Kondety, Parimala and Ortiz, Thiara and Vieira, Debra},
  booktitle={2020 53rd Annual IEEE/ACM International Symposium on Microarchitecture (MICRO)}, 
  title={Coordinated Priority-aware Charging of Distributed Batteries in Oversubscribed Data Centers}, 
  year={2020},
  pages={839-851},
  doi={10.1109/MICRO50266.2020.00073}
}

@misc{meta_orv3_grand_teton_ocp,
  title        = {OCP Summit 2022: Open hardware for AI infrastructure (Grand Teton / ORV3 rack and power)},
  author       = {{Bjorlin, Alexis}},
  url = {https://engineering.fb.com/2022/10/18/open-source/ocp-summit-2022-grand-teton},
  year         = {2022},
}

@article{safari25efficiency,
  title={A Systematic Review of Energy Efficiency Metrics for Optimizing Cloud Data Center Operations and Management},
  author={Safari, Ashkan and Sorouri, Hoda and Rahimi, Afshin and Oshnoei, Arman},
  journal={Electronics},
  volume={14},
  number={11},
  pages={2214},
  year={2025},
  publisher={MDPI}
}

@misc{microsoft_dublin_ups_grid,
  title        = {Microsoft datacenter batteries to support growth of renewables on the power grid},
  author       = {{Roach, John}},
  url = {https://news.microsoft.com/source/features/sustainability/ireland-wind-farm-datacenter-ups},
  year         = {2022}
}

@misc{eaton_microsoft_energy_aware_ups,
  title        = {Eaton and Microsoft's EnergyAware UPS technology pilot project},
  author       = {{Eaton}},
  url = {https://www.eaton.com/us/en-us/products/backup-power-ups-surge-it-power-distribution/backup-power-ups/dual-purpose-ups-technology.html}
}

@misc{nvidia_gb300_power,
  title        = {How New GB300 NVL72 Features Provide Steady Power for AI},
  author       = {Dimitrov, Rouslan and Petty, Harry and Srivastava, Neeraj and Blake, Mathias},
  url = {https://developer.nvidia.com/blog/how-new-gb300-nvl72-features-provide-steady-power-for-ai},
  year         = {2025}
}

@misc{nvidia_nvml,
  title        = {NVML API Reference Guide - GPU Deployment and Management Documentation},
  author       = {{NVIDIA}},
  url = {https://docs.nvidia.com/deploy/nvml-api/group__nvmlDeviceQueries.html#group__nvmlDeviceQueries_1gf754f109beca3a4a8c8c1cd650d7d66c},
  year         = {2025}
}

@article{aging_aware_liion_review,
title = {Aging aware operation of lithium-ion battery energy storage systems: A review},
journal = {Journal of Energy Storage},
volume = {55},
pages = {105634},
year = {2022},
issn = {2352-152X},
doi = {https://doi.org/10.1016/j.est.2022.105634},
author = {Nils Collath and Benedikt Tepe and Stefan Englberger and Andreas Jossen and Holger Hesse},
}

@article{crawford_grid_duty_cycles,
title = {Lifecycle comparison of selected Li-ion battery chemistries under grid and electric vehicle duty cycle combinations},
journal = {Journal of Power Sources},
volume = {380},
pages = {185-193},
year = {2018},
issn = {0378-7753},
doi = {https://doi.org/10.1016/j.jpowsour.2018.01.080},
author = {Alasdair J. Crawford and Qian Huang and Michael C.W. Kintner-Meyer and Ji-Guang Zhang and David M. Reed and Vincent L. Sprenkle and Vilayanur V. Viswanathan and Daiwon Choi},
}

@misc{ieee_spectrum_supercaps_ai,
  title        = {Will Supercapacitors Come to AI’s Rescue? Power bursts in large AI workloads can threaten to overwhelm the grid},
  author       = {Dina Genkina},
  url = {https://spectrum.ieee.org/supercapacitor-2671883490},
  year         = {2025}
}

@misc{eaton_supercapacitor_ai,
  title        = {Supercapacitors in AI Data Centers},
  author       = {{Eaton}},
  url = {https://www.eaton.com/us/en-us/products/electronic-components/infographics/supercaps-in-ai-datacenters.html},
}

@misc{ul9540a,
  title        = {UL 9540A Test Method for Battery Energy Storage Systems (BESS)},
  author       = {{UL Solutions}},
  url = {https://www.ul.com/services/ul-9540a-test-method},
}

@article{harmonics_review_2025,
  author  = {Kaur, Jagdeep and Bath, Sarbjeet Kaur},
  title   = {Harmonic distortion in power systems due to electronic control and renewable energy integration: a comprehensive review},
  journal = {Discover Electronics},
  volume  = {2},
  number  = {1},
  pages   = {67},
  year    = {2025},
  doi     = {10.1007/s44291-025-00111-9},
  issn    = {2948-1600}
}

@misc{iec_mvdc_overview,
  title        = {Medium voltage DC (MVDC) grids for anall-electric society},
  author       = {{IEC}},
  url = {https://www.iec.ch/basecamp/medium-voltage-dc-mvdc-grids-all-electric-society},
  year         = {2025},
}

@article{wang2025providing,
  title={Providing load flexibility by reshaping power profiles of large language model workloads},
  author={Wang, Yi and Guo, Qinglai and Chen, Min},
  journal={Advances in Applied Energy},
  pages={100232},
  year={2025},
  publisher={Elsevier}
}

@article{gu2304energy,
  author        = {Diandian Gu and Xintong Xie and Gang Huang and Xin Jin and Xuanzhe Liu},
  title         = {Energy-Efficient GPU Clusters Scheduling for Deep Learning},
  journal       = {arXiv preprint arXiv:2304.06381},
  year          = {2023},
  archivePrefix = {arXiv},
  eprint        = {2304.06381},
  primaryClass  = {cs.DC},
  url           = {https://arxiv.org/abs/2304.06381}
}

@article{al2025instability,
  author        = {Dlzar Al Kez and Aoife Foley},
  title         = {Instability Risks from Programmable AI Load Ramping in Low-Inertia Grids},
  year          = {2025},
  month         = jul,
  journal       = {SSRN Electronic Journal},
  doi           = {10.2139/ssrn.5370875},
  url           = {https://papers.ssrn.com/sol3/papers.cfm?abstract_id=5370875}
}

@ARTICLE{11230052,
  author={Farzan, Farnaz and Mahani, Khash and Farzan, Farbod and Masiello, Ralph and Brown, William},
  journal={IEEE Energy Sustainability Magazine}, 
  title={Long-Duration Energy Storage: Planning and Operation to Enhance Power Grid Sustainability}, 
  year={2025},
  volume={1},
  number={3},
  pages={41-51},
  doi={10.1109/ESM.2025.3606162}}

@techreport{norris2025rethinking,
  title        = {Rethinking Load Growth: Assessing the Potential for Integration of Large Flexible Loads in US Power Systems},
  author       = {Norris, Tyler H. and Profeta, Timothy H. and Patino-Echeverri, Dalia and Cowie-Haskell, Adam},
  institution  = {Nicholas Institute for Energy, Environment \& Sustainability, Duke University},
  year         = {2025},
  type         = {Report},
  address      = {Durham, NC},
  url          = {https://hdl.handle.net/10161/32077},
  month        = feb,
  note         = {Report NI R 25-01}
}

@online{Roy2025BigTechEnergy,
  author       = {Roy, Binita},
  title        = {New Power Players: How Big Tech Firms Are Disrupting Energy Markets},
  year         = {2025},
  date         = {2025-09-29},
  url          = {https://www.uh.edu/energy/news/stories/2025/ow_big_tech_firms_are_disrupting_energy_hmarkets.php},
  organization = {University of Houston Energy},
  note         = {Accessed: 2026-01-24},
}

@article{White2020RepurposedSLBFR,
  title   = {Repurposed electric vehicle battery performance in second-life electricity grid frequency regulation service},
  author  = {White, C. and Thompson, B. and Swan, L. G.},
  journal = {Journal of Energy Storage},
  volume  = {28},
  pages   = {101278},
  year    = {2020},
  doi     = {10.1016/j.est.2020.101278}
}

@article{majeed2022impact,
  title={Impact of reverse power flow on distributed transformers in a solar-photovoltaic-integrated low-voltage network},
  author={Majeed, Issah Babatunde and Nwulu, Nnamdi I},
  journal={Energies},
  volume={15},
  number={23},
  pages={9238},
  year={2022},
  publisher={MDPI}
}

@article{colangelo2025ai,
  title={AI data centres as grid-interactive assets},
  author={Colangelo, Philip and Coskun, Ayse K and Megrue, Jack and Roberts, Ciaran and Sengupta, Shayan and Sivaram, Varun and Tiao, Ethan and Vijaykar, Aroon and Williams, Chris and Wilson, Daniel C and others},
  journal={Nature Energy},
  pages={1--8},
  year={2025},
  publisher={Nature Publishing Group UK London}
}

@article{paananen2021grid,
  title={Grid-interactive data centers: enabling decarbonization and system stability},
  author={Paananen, Janne and Nasr, Ehsan},
  journal={Dublin, Ireland},
  year={2021}
}

@article{takci2025data,
  title={Data centres as a source of flexibility for power systems},
  author={Takci, Mehmet T{\"u}rker and Qadrdan, Meysam and Summers, Jon and Gustafsson, Jonas},
  journal={Energy Reports},
  volume={13},
  pages={3661--3671},
  year={2025},
  publisher={Elsevier}
}

@ARTICLE{7879307,
  author={Xu, Bolun and Wang, Yishen and Dvorkin, Yury and Fernández-Blanco, Ricardo and Silva-Monroy, Cesar A. and Watson, Jean-Paul and Kirschen, Daniel S.},
  journal={IEEE Transactions on Power Systems}, 
  title={Scalable Planning for Energy Storage in Energy and Reserve Markets}, 
  year={2017},
  volume={32},
  number={6},
  pages={4515-4527},
  doi={10.1109/TPWRS.2017.2682790}}

@ARTICLE{8805438,
  author={Padmanabhan, Nitin and Ahmed, Mohamed and Bhattacharya, Kankar},
  journal={IEEE Transactions on Power Systems}, 
  title={Battery Energy Storage Systems in Energy and Reserve Markets}, 
  year={2020},
  volume={35},
  number={1},
  pages={215-226},
  doi={10.1109/TPWRS.2019.2936131}}

@article{liang2026gpu,
  title={GPU-to-Grid: Voltage Regulation via GPU Utilization Control},
  author={Liang, Zhirui and Chung, Jae-Won and Chowdhury, Mosharaf and Chen, Jiasi and Dvorkin, Vladimir},
  journal={arXiv preprint arXiv:2602.05116},
  year={2026}
}

@article{azizi2026strengthening,
  title={Strengthening Data Center Operations Using Grid-Forming Battery Energy Storage as a Line-Interactive Uninterruptible Power Supply},
  author={Azizi, Arian and Morovati, Samaneh and Zamani, Amin and Piruzza, Joshua and Guo, David and Liu, Zhuoning and Taimela, Pasi},
  journal={International Journal of Electrical Power \& Energy Systems},
  volume={175},
  pages={111638},
  year={2026},
  publisher={Elsevier}
}

@article{chung2025ml,
  title={The ML. ENERGY benchmark: Toward automated inference energy measurement and optimization},
  author={Chung, Jae-Won and Ma, Jeff J and Wu, Ruofan and Liu, Jiachen and Kweon, Oh Jun and Xia, Yuxuan and Wu, Zhiyu and Chowdhury, Mosharaf},
  journal={arXiv preprint arXiv:2505.06371},
  year={2025}
}

@article{chung2026joules,
  title={Where Do the Joules Go? Diagnosing Inference Energy Consumption},
  author={Chung, Jae-Won and Wu, Ruofan and Ma, Jeff J and Chowdhury, Mosharaf},
  journal={arXiv preprint arXiv:2601.22076},
  year={2026}
}

@article{wu2026kareus,
  title={Kareus: Joint Reduction of Dynamic and Static Energy in Large Model Training},
  author={Wu, Ruofan and Chung, Jae-Won and Chowdhury, Mosharaf},
  journal={arXiv preprint arXiv:2601.17654},
  year={2026}
}

@inproceedings{chung2024reducing,
  title={Reducing energy bloat in large model training},
  author={Chung, Jae-Won and Gu, Yile and Jang, Insu and Meng, Luoxi and Bansal, Nikhil and Chowdhury, Mosharaf},
  booktitle={Proceedings of the ACM SIGOPS 30th Symposium on Operating Systems Principles},
  pages={144--159},
  year={2024}
}

@article{conto2025texas,
  title={Texas Loads Ride Toward Grid Stability: Voltage Ride Through of Large Power Electronic Loads},
  author={Conto, Jos{\'e} and Cheng, Yunzhi and Rose, Jonathan and Schmall, John},
  journal={IEEE Power and Energy Magazine},
  volume={23},
  number={5},
  pages={56--67},
  year={2025},
  publisher={IEEE}
}

@article{zahedi2026best,
  title={Best Practices for Large Load Interconnections: A North American Perspective on Data Centers},
  author={Zahedi, Rafi and Zamani, Amin and Anilkumar, Rahul},
  journal={arXiv preprint arXiv:2601.12686},
  year={2026}
}

@article{li2016fault,
  title={Fault ride-through of renewable energy conversion systems during voltage recovery},
  author={Li, Ruiqi and Geng, Hua and Yang, Geng},
  journal={Journal of Modern Power Systems and Clean Energy},
  volume={4},
  number={1},
  pages={28--39},
  year={2016},
  publisher={SGEPRI}
}

@article{cheng2011voltage,
  title={Voltage-profile-based approach for developing collection system aggregated models for wind generation resources for grid voltage ride-through studies},
  author={Cheng, Y and Sahni, M and Conto, J and Huang, S-H and Schmall, J},
  journal={IET Renewable Power Generation},
  volume={5},
  number={5},
  pages={332--346},
  year={2011},
  publisher={IET}
}

@article{pant2011introduction,
  title={An introduction to the life cycle assessment (LCA) of bioelectrochemical systems (BES) for sustainable energy and product generation: relevance and key aspects},
  author={Pant, Deepak and Singh, Anoop and Van Bogaert, Gilbert and Gallego, Yolanda Alvarez and Diels, Ludo and Vanbroekhoven, Karolien},
  journal={Renewable and Sustainable Energy Reviews},
  volume={15},
  number={2},
  pages={1305--1313},
  year={2011},
  publisher={Elsevier}
}

@article{kaiser2007optimized,
  title={Optimized battery-management system to improve storage lifetime in renewable energy systems},
  author={Kaiser, Rudi},
  journal={Journal of Power Sources},
  volume={168},
  number={1},
  pages={58--65},
  year={2007},
  publisher={Elsevier}
}

@article{zhang2019optimal,
  title={Optimal whole-life-cycle planning of battery energy storage for multi-functional services in power systems},
  author={Zhang, Yongxi and Xu, Yan and Yang, Hongming and Dong, Zhao Yang and Zhang, Rui},
  journal={IEEE Transactions on Sustainable Energy},
  volume={11},
  number={4},
  pages={2077--2086},
  year={2019},
  publisher={IEEE}
}

@article{wang2026life,
  title={Life-cycle performance analysis of a building integrated energy system considering equipment performance degradation},
  author={Wang, Jiangjiang and Ye, Shaoming and Wu, Boling and Liu, Boxiang},
  journal={Energy Conversion and Management},
  volume={347},
  pages={120593},
  year={2026},
  publisher={Elsevier}
}

@article{kamali2016life,
  title={Life cycle performance of modular buildings: A critical review},
  author={Kamali, Mohammad and Hewage, Kasun},
  journal={Renewable and sustainable energy reviews},
  volume={62},
  pages={1171--1183},
  year={2016},
  publisher={Elsevier}
}

@article{omar2014lithium,
  title={Lithium iron phosphate based battery--Assessment of the aging parameters and development of cycle life model},
  author={Omar, Noshin and Monem, Mohamed Abdel and Firouz, Yousef and Salminen, Justin and Smekens, Jelle and Hegazy, Omar and Gaulous, Hamid and Mulder, Grietus and Van den Bossche, Peter and Coosemans, Thierry and others},
  journal={Applied Energy},
  volume={113},
  pages={1575--1585},
  year={2014},
  publisher={Elsevier}
}

@article{majeau2011life,
  title={Life cycle environmental assessment of lithium-ion and nickel metal hydride batteries for plug-in hybrid and battery electric vehicles},
  author={Majeau-Bettez, Guillaume and Hawkins, Troy R and Str{\o}mman, Anders Hammer},
  journal={Environmental science \& technology},
  volume={45},
  number={10},
  pages={4548--4554},
  year={2011},
  publisher={ACS Publications}
}

@article{ding2019automotive,
  title={Automotive Li-ion batteries: current status and future perspectives},
  author={Ding, Yuanli and Cano, Zachary P and Yu, Aiping and Lu, Jun and Chen, Zhongwei},
  journal={Electrochemical Energy Reviews},
  volume={2},
  number={1},
  pages={1--28},
  year={2019},
  publisher={Springer}
}

@inproceedings{irion2025optimizing,
  title={Optimizing Microgrid Composition for Sustainable Data Centers},
  author={Irion, Julius and Wiesner, Philipp and Bader, Jonathan and Kao, Odej},
  booktitle={Proceedings of the SC'25 Workshops of the International Conference for High Performance Computing, Networking, Storage and Analysis},
  pages={1990--1996},
  year={2025}
}

@article{zhang2025unlocking,
  title={Unlocking the flexibilities of data centers for smart grid services: Optimal dispatch and design of energy storage systems under progressive loading},
  author={Zhang, Yingbo and Tang, Hong and Li, Hangxin and Wang, Shengwei},
  journal={Energy},
  volume={316},
  pages={134511},
  year={2025},
  publisher={Elsevier}
}

@inproceedings{mughees2025short,
  title={Short-Term Load Forecasting for AI-Data Center},
  author={Mughees, Mariam and Li, Yuzhuo and Chen, Yize and Li, Yunwei Ryan},
  booktitle={2025 IEEE Power \& Energy Society General Meeting (PESGM)},
  pages={1--5},
  year={2025},
  organization={IEEE}
}

@article{jiang2025hyperload,
  title={HyperLoad: A Cross-Modality Enhanced Large Language Model-Based Framework for Green Data Center Cooling Load Prediction},
  author={Jiang, Haoyu and Qu, Boan and Zhu, Junjie and Zeng, Fanjie and Lin, Xiaojie and Zhong, Wei},
  journal={arXiv preprint arXiv:2512.19114},
  year={2025}
}

@article{mo2025learning,
  title={Learning from imbalance: Cross-server power prediction in large data centers via domain adaptation regression},
  author={Mo, Ruichao and Lin, Weiwei and Liu, Guozhi and Liu, Haolin and He, Ligang},
  journal={Expert Systems with Applications},
  volume={287},
  pages={127845},
  year={2025},
  publisher={Elsevier}
}

@article{8571261,
  author={Rothmund, Daniel and Guillod, Thomas and Bortis, Dominik and Kolar, Johann W.},
  journal={IEEE Journal of Emerging and Selected Topics in Power Electronics}, 
  title={99\% Efficient 10 kV SiC-Based 7 kV/400 V DC Transformer for Future Data Centers}, 
  year={2019},
  volume={7},
  number={2},
  pages={753-767},
  doi={10.1109/JESTPE.2018.2886139}}

@article{cao2025transforming,
  title={Transforming future data center operations and management via physical AI},
  author={Cao, Zhiwei and Li, Minghao and Lin, Feng and Jia, Jimin and Wen, Yonggang and Yin, Jianxiong and See, Simon},
  journal={arXiv preprint arXiv:2504.04982},
  year={2025}
}

@article{li2025phythesis,
  title={Phythesis: Physics-Guided Evolutionary Scene Synthesis for Energy-Efficient Data Center Design via LLMs},
  author={Li, Minghao and Wang, Ruihang and Tan, Rui and Wen, Yonggang},
  journal={arXiv preprint arXiv:2512.10611},
  year={2025}
}

@article{crozier2025potential,
  title={The potential of data center energy demand to provide grid flexibility},
  author={Crozier, Constance and Liska, Matthew},
  journal={Current Sustainable/Renewable Energy Reports},
  volume={12},
  number={1},
  pages={12},
  year={2025},
  publisher={Springer}
}

@misc{FrotscherRaveTeNyenhuisUpadhyay2021RPF,
  author       = {Frotscher, Rainer and Rave, Martin and teNyenhuis, Ed and Upadhyay, Parag},
  title        = {Reverse Power Flow Impact on Transformers},
  howpublished = {IEEE PES Transformers Committee Spring 2021 Meeting -- Tutorial / Technical Presentation},
  year         = {2021},
  month        = apr,
  day          = {29},
  url          = {https://grouper.ieee.org/groups/transformers/subcommittees/standardsc/C57.133/F24-C57.133-ReversePowerFlowTutorial-%20teNyenhuis.pdf},
  note         = {Slides}
}

@article{li2025ups_datacenter,
  title   = {Coordinating Multiple UPS Batteries in Datacenter for Load Flexibility and Cost Reduction},
  author  = {Lunlong Li and Pu Yang and Yi Ju and Ziqi Hu and Zhe Wang},
  journal = {SSRN Electronic Journal},
  year    = {2025},
  doi     = {10.2139/ssrn.5647230},
  url     = {https://ssrn.com/abstract=5647230}
}

\end{document}